\documentclass[11pt,a4paper]{article}
\pdfoutput=1
\usepackage{jheppub}

\usepackage{multirow, graphicx,amssymb,url,mathrsfs,amsmath}
\usepackage{wrapfig,boxedminipage,subfigure,epsfig}
\usepackage{amsxtra,amstext,latexsym,dsfont,amsfonts}
\usepackage{color}
\usepackage[dvipsnames]{xcolor}
\usepackage{float}
\usepackage{slashed}
\usepackage{calligra}
\DeclareFontShape{T1}{calligra}{m}{n}{<->s*[2.2]callig15}{}
\DeclareMathAlphabet{\mathcalligra}{T1}{calligra}{m}{n}
\usepackage{tikz} 
\usepackage[compat=1.1.0]{tikz-feynman}
\usepackage{feynmf}

\newcommand{\be}{\begin{equation}}
\newcommand{\ee}{\end{equation}}
\newcommand{\bea}{\begin{eqnarray}}
\newcommand{\eea}{\end{eqnarray}}

\newcommand{\half}{\frac{1}{2}}

\title{Holographic teleportation in higher dimensions}

\author[]{Byoungjoon Ahn,}
\author[]{Yongjun Ahn,}
\author[]{Sang-Eon Bak,}
\author[]{Viktor Jahnke,}
\author[]{Keun-Young Kim}

\affiliation[]{School of Physics and Chemistry, Gwangju Institute of Science and Technology, 123 Cheomdan-gwagiro, Gwangju 61005, Korea}

\emailAdd{bjahn123@gist.ac.kr}
\emailAdd{yongjunahn619@gmail.com}
\emailAdd{sangeonbak@gm.gist.ac.kr}
\emailAdd{viktorjahnke@gist.ac.kr}
\emailAdd{fortoe@gist.ac.kr}

\abstract{We study higher-dimensional traversable wormholes in the context of Rindler-AdS/CFT. The hyperbolic slicing of a pure AdS geometry can be thought of as a topological black hole that is dual to a conformal field theory in the hyperbolic space. The maximally extended geometry contains two exterior regions (the Rindler wedges of AdS) which are connected by a wormhole. We show that this wormhole can be made traversable by a double trace deformation that violates the average null energy condition (ANEC) in the bulk. We find an analytic formula for the ANEC violation that generalizes Gao-Jafferis-Wall result to higher-dimensional cases, and we show that the same result can be obtained using the eikonal approximation. We show that the bound on the amount of information that can be transferred through the wormhole quickly reduces as we increase the dimensionality of spacetime. We also compute a two-sided commutator that diagnoses traversability and show that, under certain conditions, the information that is transferred through the wormhole propagates with butterfly speed $v_B = \frac{1}{d-1}$.
}

\begin{document}
\maketitle

\section{Introduction}

In the context of gauge-gravity duality \cite{Maldacena:1997re,Witten:1998qj,Gubser:1998bc}, an asymptotically AdS two-sided black hole geometry is dual to two copies of the corresponding boundary theory entangled in a thermofield double state \cite{Maldacena:2001kr}. The geometry contains a wormhole connecting the two asymptotic regions where the boundary theories live. The wormhole is not traversable, which is consistent with the fact that the boundary theories do not interact with each other.

In general, the non-traversability of wormholes is a consequence of the Average Null Energy Condition (ANEC), which basically states that the stress energy tensor integrated along a complete achronal null geodesic is always non-negative
\be \label{eq-ANEC-general}
\int T_{\mu \nu} k^{\mu} k^{\nu} d\lambda \geq 0\,,
\ee
where $k^{\mu}$ is a tangent vector and $\lambda$ is an affine parameter. 
\eqref{eq-ANEC-general} establishes that the Average Null Energy (ANE) can be used as a diagnose of traversability.




Gao, Jafferis and Wall(GJW) realized that ANEC can be violated and the wormhole can be made traversable in a two-sided BTZ black hole by coupling the two boundary theories with a relevant perturbation of the form
\be \label{eq-deformation}
\delta H = \int dt d{\bf x}\, h(t,{\bf x})\mathcal{O}_{L}(-t,{\bf x}) \mathcal{O}_R(t,{\bf x})\,,
\ee
where $\mathcal{O}$ is a scalar operator dual to a bulk scalar field $\phi$\,\cite{Gao_2017}.
They work in the context of a semiclassical approximation, in which the gravitational field is treated classically, while the matter fields are treated quantum mechanically. In this context, one writes the Einstein's equation by replacing the matter stress energy tensor by its expectation value in a given state
\be
G_{\mu \nu} = 8 \pi G_N \langle T_{\mu \nu}\rangle\,,
\ee
where $G_{\mu \nu}$ is the Einstein's tensor. GJW showed that the double trace deformation can be chosen in such a way that makes the expectation value of the quantum matter stress tensor violate the ANEC \eqref{eq-ANEC-general}. The physical picture is that the boundary deformation introduces a field excitation in the bulk that has negative energy. The backreaction of this negative energy is given in terms of a negative-energy shock wave that causes a time advance for the geodesic, as opposed to the usual time delay caused by positive-energy shock waves. This allows us to transfer information through the wormhole, as shown in Fig.~\ref{fig-info-tranfer}.
In the dual field theory description, the traversability of the wormhole is related to a teleportation protocol \cite{Maldacena_2017,gao2021traversable}. The above-mentioned results lead to several interesting developments  \cite{almheiri2018escaping,Caceres_2018,Bak_2018,Bak_2019, Bak_2019_2,Couch_2020,Freivogel_2020,fallows2020making,Geng2020kxhx,nosaka2020chaos,levine2020seeing, Fu_2019, Marolf_2019, balushi2021traversability, emparan2020multimouth,Bao_2018,Hirano_2019,Garc_a_Garc_a_2019,numasawa2020coupled}, including a proposal for studying quantum gravity experimentally \cite{brown2019quantum,nezami2021quantum}.

Most of the works involving traversable wormholes by a double trace deformation only deal with lower dimensional cases, like black holes in 2D or 3D gravity, while in more realistic experimental setups one usually expects higher-dimensional systems. In fact, due to technical problems, the case of higher-dimensional black holes is more complicated, and it has not been explored in detail\footnote{See, however, \cite{Freivogel_2020}.}. In this work, we fill this gap in the literature by studying traversable wormholes in the context of Rindler-AdS/CFT. We generalize GJW results to the case of a Rindler-AdS$_{d+1}$ ($d \ge 2$) geometry and show that the same results can be obtained using the eikonal approximation, as done in \cite{Maldacena_2017} for a 2D gravity theory.

Another motivation for our work comes from the existence of a no-go theorem regarding eternal traversable wormholes in higher dimensions \cite{Freivogel_2019}. In fact, by considering a pair of unentangled CFTs, assuming Poincare invariance in the boundary directions, and using Weyl invariant matter fields, the authors of \cite{Freivogel_2019} proved the non-existence of eternal semi-classical traversable wormholes in spacetime dimensions higher than two.  In this paper we point out that it is possible to evade this no-go theorem and explicitly construct higher dimensional examples of traversable wormholes if we assume: (i) non-eternal traversable wormholes, (ii) two entangled copies of the CFT, and (iii) matter fields without Weyl symmetry.



This work is organized as follows. In Sec.~\ref{sec-gravity}, we review our holographic setup. In Sec.~\ref{sec-ANEC}, we derive analytic formulas for the ANEC violation, discuss backreaction effects and derive semi-analytic formulas for a two-sided correlator that diagnoses traversability. In Sec.~\ref{sec-bound}, we derive a parametric bound on the amount of information that can be transferred through the wormhole and discuss the dependence on the dimensionality of the spacetime. In Sec.~\ref{sec-deltaE}, we compute the change of energy and entropy that result from the double trace deformation. We discuss our results in Sec.~\ref{sec-disc} and relegate some technical details to Appendix \ref{app:A}.

\section{Gravity set-up} \label{sec-gravity}
We work in the context of Einstein gravity
\be
S=\frac{1}{16 \pi G_N} \int d^{d+1}x
\left[ R + \frac{d(d-1)}{\ell^2} \right]\,,
\ee
and we consider a Rindler-AdS solution, which can be constructed as follows.
We start with a pure $AdS_{d+1}$ geometry, which can be defined as the universal cover of the hyperboloid
\be
-T_1^2-T_2^2+X_1^2+\cdots +X_d^2=-\ell^2\,,
\ee
embedded in a space with ambient metric given by
\be
ds^2=-dT_1^2-dT_2^2+dX_1^2+ \cdots + dX_d^2\,.
\label{eq-metric-emb}
\ee
Here, $\ell$ denotes the AdS length scale. In what follows, we set $\ell=1$ for simplicity. We then parametrize the embedding coordinates as follows
\begin{align}
\begin{split}
T_1&=\sqrt{r^2-1} \sinh t\,,\\
T_2&= r \cosh \chi,\\
X_d&= \sqrt{r^2-1} \cosh t,\\
X_1^2+\cdots +X_{d-1}^2 &= r^2 \sinh^2 \chi\,,
\label{eq-emb-rindler}
\end{split}
\end{align}
where $t \in (-\infty, \infty)$, and $r, \chi \in [0,\infty)$. In terms of these coordinates, the metric becomes
\be
ds^2=-\left(r^2-1\right)dt^2+\frac{dr^2}{r^2-1}+r^2 d\mathbb{H}_{d-1}^2\,,
\label{eq-metric-rindler}
\ee
where $d\mathbb{H}_{d-1}^2=d\chi^2+\sinh^2\chi\, d \Omega_{d-2}^2$ denotes the unity metric in a $(d-1)-$dimensional hyperbolic space, $\mathbb{H}_{d-1}$, with $d \Omega_{d-2}^2$ being the unity metric on the $(d-2)-$sphere. The AdS boundary is located at $r=\infty$, and the geometry has a horizon at $r=1$, which leads to a non-zero Hawking temperature given by $T=\frac{1}{\beta}=\frac{1}{2 \pi}$.

\begin{figure}
\centering
\includegraphics[width=6cm]{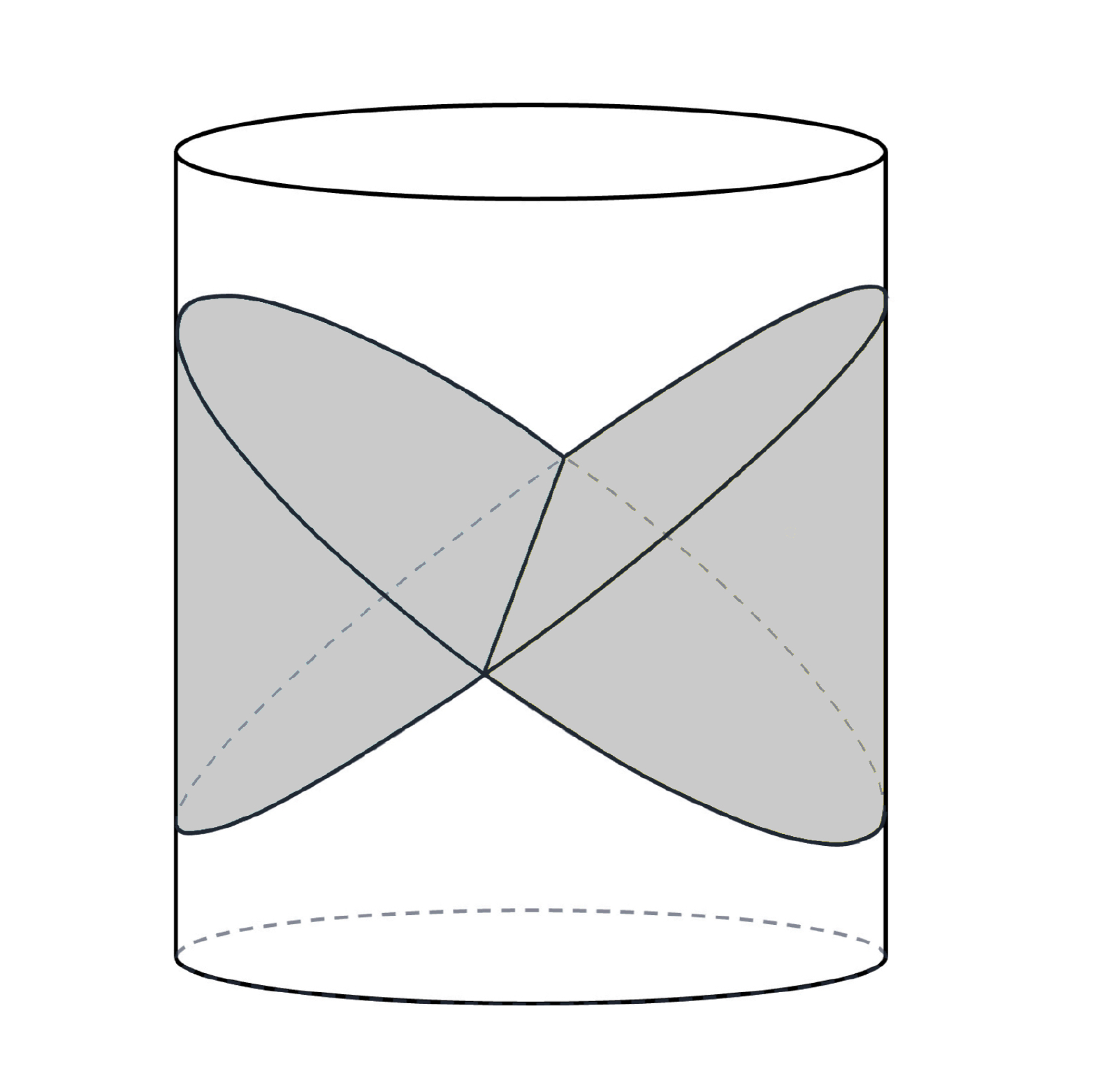}
\caption{Rindler wedges of global AdS (solid cylinder)}
\label{fig-rindlerwedge}
\end{figure}

The coordinates $(t,r,{\bf x})$, where ${\bf x} \in \mathbb{H}_{d-1}$, describe an accelerating observer in AdS, and they only cover a subregion of the spacetime known as the Rindler wedge of AdS,\footnote{In the literature, the solution (\ref{eq-metric-rindler}) is also referred to as a `topological' hyperbolic black hole \cite{Emparan:1999gf}.} which is shown in light gray in Figure \ref{fig-rindlerwedge}. In this particular hyperbolic foliation of AdS, the dual boundary theory is a CFT$_d$ living in $\mathbb{R} \times \mathbb{H}_{d-1}$.

\begin{figure}
\centering

\begin{tikzpicture}[scale=1.5]
\draw [thick]  (0,0) -- (0,3);
\draw [thick]  (3,0) -- (3,3);
\draw [thick,dashed]  (0,0) -- (3,3);
\draw [thick,dashed]  (0,3) -- (3,0);
\draw [thick,decorate,decoration={zigzag,segment length=1.5mm, amplitude=0.3mm}] (0,3) .. controls (.75,2.85) 
and (2.25,2.85) .. (3,3);
\draw [thick,decorate,decoration={zigzag,segment length=1.5mm,amplitude=.3mm}]  (0,0) .. controls (.75,.15) and (2.25,.15) .. (3,0);

\draw[thick,<->] (1,2.2) -- (1.5,1.7) -- (2,2.2);

\node[scale=0.8, align=center] at (1.5,2.65) {Future Interior};
\node[scale=0.8,align=center] at (1.5,.55) {Past Interior};
\node[scale=0.8,align=center] at (0.6,1.6) {Left\\ Exterior};
\node[scale=0.8,align=center] at (2.4,1.6) {Right\\ Exterior};
\end{tikzpicture}
\vspace{0.1cm}
\put(-142,50){\rotatebox{90}{\small $r = \infty$}}
\put(-0,80){\rotatebox{-90}{\small $r = \infty$}}
\put(-75,-5){\small $r = 0$}
\put(-75,130){\small $r = 0$}
\put(-96,95){\small $V$}
\put(-46,95){\small $U$}

\caption{ \small Penrose diagram for two-sided black holes with asymptotically AdS geometry.}
\label{fig-Penrose}
\end{figure}
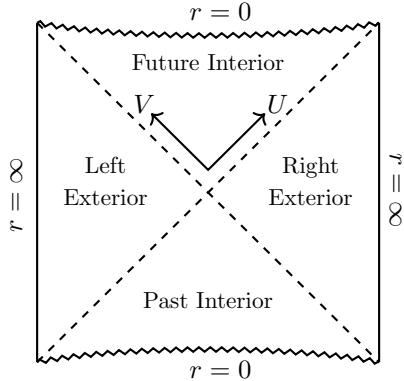

The full spacetime can be described either in Kruskal-Szekeres coordinates, or in global coordinates. Let us start by introducing Kruskal-Szekeres coordinates $(U,V)$ as
\begin{align}
\begin{split}
V&=+e^{r_*-t}\,,\,\,\,\,U=-e^{r_*+t}\,,\,\,\,\,\text{left wedge}\\
V&=-e^{r_*-t}\,,\,\,\,\,U=+e^{r_*+t}\,,\,\,\,\,\text{right wedge}\\
V&=+e^{r_*-t}\,,\,\,\,\,U=+e^{r_*+t}\,,\,\,\,\,\text{future interior}\\
V&=-e^{r_*-t}\,,\,\,\,\,U=-e^{r_*+t}\,,\,\,\,\,\text{past interior}
\end{split}   
\end{align}
where the tortoise coordinate is defined as
\be
r_*=\int^r \frac{dr'}{r'^2-1}= \log \left( \frac{r-1}{r+1}\right)^{1/2}\,.
\ee
In terms of Kruskal-Szekeres coordinates, the metric becomes
\be \label{eq-metric-kruskal}
ds^2=-\frac{4 dU \,dV}{(1+U V)^2}+\left( \frac{1-U V}{1+ U V} \right)^2 d\mathbb{H}_{d-1}^2\,.
\ee
The above result can also be obtained by substituting the following embedding coordinates
\begin{align}
    \begin{split}
T_1&=\frac{U+V}{1+U V} \,,\\
T_2&= \frac{1-U V}{1+U V} \cosh \chi,\\
X_d&= \frac{U-V}{1+U V},\\
X_1^2+\cdots +X_{d-1}^2 &= \left( \frac{1-U V}{1+U V} \right)^2 \sinh^2 \chi\,,
\label{eq-emb-kruskal}        
    \end{split}
\end{align}
into the embedding space metric (\ref{eq-metric-emb}). In Kruskal-Szekeres coordinates $(U,V,{\bf x})$, the AdS boundary is located at $UV=-1$, and there is a coordinate singularity at $UV=1$. The geometry contains two horizons, which are located at $U=0$ and $V=0$. Figure \ref{fig-Penrose} shows the corresponding Penrose diagram. The two exterior regions correspond to the two Rindler wedges of AdS. 

Each Rindler wedge is described by a CFT living in $\mathbb{R} \times \mathbb{H}_{d-1}$. We denote these hyperbolic space CFTs as CFT$_L$ and CFT$_R$, where $L$ and $R$ label the left and the right boundary, respectively. The maximally extended geometry (\ref{eq-emb-kruskal}) is dual to a thermofield double state constructed by entangling the two hyperbolic space CFTs
\be \label{eq-tfd}
|\text{TFD} \rangle = Z^{-1/2} \sum_n e^{-\beta E_n/2}  | E_n \rangle_L  \otimes |E_n\rangle_R,
\ee
where $L$ and $R$ label states in CFT$_L$ and CFT$_R$, respectively, and $Z=\sum_n e^{-\beta E_n}$ is the thermal partition function at inverse temperature $\beta$. The Rindler-AdS geometry fixes $\beta=2\pi$.

The description above shows that a global AdS geometry can be seen as a maximally extended black hole-like geometry, whose boundary description is given in terms of a thermofield double state of two CFTs in hyperbolic space. This can be understood as follows. In global coordinates, one chooses the following parametrization 
\begin{align}
T_1=\sqrt{\rho^2+1}\cos \tau\,,\,\,\, T_2=\sqrt{\rho^2+1} \sin \tau\,,\,\,\, X_1^2+\cdots + X_d^2=\rho^2\,, 
\end{align}
in terms of which the metric becomes
\be
ds^2= - (1+\rho^2) d\tau^2 + \frac{d\rho^2}{1+\rho^2}+\rho^2 d\Omega_{d-1}^2\,,
\ee
where $\tau \in (-\infty,\infty),\,\, \rho \in [0,\infty)$. The AdS boundary is located at $\rho = \infty$ and the dual field theory description is given in terms of a CFT in $\mathbb{R} \times S^{d-1}$. In particular, the pure global AdS geometry describes the vacuum state $| 0 \rangle_{\text{global}}$ of such CFT.

Given a constant time slice of the geometry, we can divide the spatial boundary into two hemispheres $B_L$ and $B_R$, and decompose the Hilbert space accordingly, i.e., $\mathcal{H} = \mathcal{H}_{B_L} \otimes \mathcal{H}_{B_R}$. It turns out that the vacuum state of the CFT defined on the full sphere can be written as a thermofield double state constructed out of the states of the CFTs defined on the hemispheres. Finally, we can use conformal transformations to map the domain of dependence of both $B_L$ and $B_R$ to $\mathbb{R} \times \mathbb{H}_{d-1}$.
This implies that vacuum state $| 0 \rangle_{\text{global}}$ can be written as a thermofield double state of two hyperbolic space CFTs \cite{Czech_2012,Van_Raamsdonk_2016}
\be
| 0 \rangle_{\text{global}} = Z^{-1/2} \sum_n e^{-\pi E_n}  | E_n \rangle_L  \otimes |E_n\rangle_R\,,
\ee
where $E_L$ and $E_R$ label the energy eigenstates of CFT$_L$ and CFT$_R$. This provides the field theory explanation of why global AdS can be thought of as a maximally extended hyperbolic `black hole'.

The above discussion is related to the so-called subregion duality, which states that if we only have access to a subset of the boundary, we can only describe a subregion of the bulk geometry. In this particular example, a CFT that has support on only half of the boundary of global AdS will not describe the full bulk geometry, but only the corresponding Rindler wedge of AdS.
For a more detailed discussion about subregion duality, we refer to \cite{harlow2018tasi}.

\subsection{Bulk-boundary propagators}
In this section, we compute bulk-boundary propagators of scalar fields in the maximally extended Rindler-AdS geometry. These propagators will be important ingredients in the computation of ANEC violation in the following sections. We first compute bulk-bulk propagators between two bulk points, and then we obtain bulk-boundary propagators by taking one of these points to the boundary. 

These propagators are given in terms of geodesic distances between two points.
In embedding coordinates, the geodesic distance $d(P,P')$ between two points $P=(T_1,T_2,X_i)$ and $P'=(T_1',T_2',X_i')$ in the ambient space can be written as
\be
\cosh d(P,P')= T_1 T_1'+T_2 T_2'-\sum_{i=1}^{d}X_i X_i'\,.
\ee
The bulk-bulk propagator between $P$ and $P'$ is given by \cite{Ammon:2015wua}
\bea
&&G_{\Delta}(P;P')=c_{\Delta} \xi^{\Delta}\,{}_2F_1\left( \frac{\Delta}{2},\frac{\Delta+1}{2};\Delta+1-\frac{d}{2}; \xi^2\right)\,,\\
&&\xi \equiv \left( \cosh d(P,P') \right)^{-1}\,,\,\,\,\, c_{\Delta} \equiv \frac{\Gamma(\Delta)}{2^{\Delta+1}\pi^{d/2}\Gamma(\Delta+1-d/2)}\,, \nonumber\label{eq-cdelta}
\eea
where $\Delta$ is the conformal dimension of the scalar operator.

The bulk-boundary propagator can be computed by taking $P$ or $P'$ to the boundary. It is more convenient to do that after specifying a coordinate system. By writing the embedding coordinates in terms of Rindler-AdS coordinates $(t,r,{\bf x})$, we can show that
\be
\cosh d(P,P')= - \sqrt{r^2-1} \sqrt{r'^2-1} \cosh(t-t')+r r' \cosh d({\bf x,x'})\,,
\ee
where $d({\bf x,x'})$ is the geodesic distance between ${\bf x}$ and ${\bf x'}$ in $\mathbb{H}_{d-1}$. Here, we follow the convention of GJW and compute the bulk-boundary propagator as
\footnote{It is  also customary to define the bulk-boundary propagator as \cite{Ammon:2015wua}\\
\be
 K_{\Delta}(t,r,{\bf x};t',{\bf x'})\equiv (2\Delta-d)\lim_{r' \rightarrow \infty} G_{\Delta}(t',r',{\bf x'})\,\nonumber.
\ee 
 The factor of $2\Delta-d$ is included to guarantee that $\lim_{r' \rightarrow \infty} r'^{d-\Delta} K_{\Delta}(t,r,{\bf x};t',{\bf x'})=\delta(t-t')\delta^{d-1}({\bf x-x'})$.}
\be
K_{\Delta}(U,V,{\bf x};t',{\bf x'})\equiv \langle \phi(U,V,{\bf x}) \mathcal{O}(t',{\bf x'}) \rangle =  \lim_{r' \rightarrow \infty} r'^{\Delta} G_{\Delta}(r,t,{\bf x};r',t',{\bf x'})\,,
\ee
where $\mathcal{O}$ is a boundary operator, with conformal dimension $\Delta$, and $\phi$ is the corresponding dual bulk field.
With the above definition we obtain
\be
K_{\Delta}(r,t,{\bf x};t',{\bf x'})= c_{\Delta} \left[ - \sqrt{r^2-1} \cosh(t-t')+r \cosh d({\bf x,x'}) \right]^{-\Delta}.
\ee
For later purposes, it will be convenient to write the bulk-point $(r,t,{\bf x})$ in terms of Kruskal-Szekeres coordinates $(U,V,{\bf x})$. In this case, the propagator becomes
\be \label{eq-bbprop}
K_{\Delta}(U,V,{\bf x};t',{\bf x'}) = c_{\Delta} \left( \frac{1+UV}{V e^{t'}-Ue^{-t'}+(1-UV) \cosh d({\bf x,x'})} \right)^{\Delta}\,.
\ee
The above formula is valid when both the bulk point $ (U,V,{\bf x}) $ and the boundary point $ (t',{\bf x'})$ are in the right exterior region (right Rindler wedge). Later, we will also need the formula for the case in which the boundary point is on the left asymptotic boundary. This formula can be simply obtained by replacing $ t \rightarrow -t + i \pi $ in (\ref{eq-bbprop}).

\section{ANEC violation and traversable wormhole} \label{sec-ANEC}
Let us now review how exactly the violation of the ANEC leads to a traversable wormhole in a Rindler-AdS$_{d+1}$ geometry. We start by considering the linearized Einstein's equation in Kruskal coordinates
\begin{equation} \label{eq-EOM}
    \frac{d-1}{2\ell^2}\big(h_{UU}+\partial_{U} (U h_{UU})\big)+\frac{1}{2\ell^2} \partial_{U}^{2} h_{\chi \chi} =8\pi G_N T_{UU}\,,
\end{equation}
where we denote the fluctuations as $\delta g_{\mu \nu} = h_{\mu \nu}$. Integrating \eqref{eq-EOM} with respect to $U$ with $\ell=1$, we obtain
\be \label{eq-hUUintegral}
 \frac{d-1}{2}\int  h_{UU} dU = 8 \pi G_N \int  T_{UU} dU\,.
\ee
A null ray which originates from the past infinity to future infinity along the horizon $(V=0)$ undergoes a shift in the $V$ direction by
\begin{equation} \label{eq-DELTAV}
\Delta V (U)= -\frac{1}{2g_{UV}(0)} \int_{-\infty}^{U}  h_{UU} dU\,,
\end{equation}
where $g_{UV}(0)=-2$ for the Rindler-AdS$_{d+1}$ geometry.
Combining \eqref{eq-hUUintegral} and \eqref{eq-DELTAV}, we find
\be \label{eq-deltav-ANEC}
\Delta V = \frac{4 \pi G_N}{d-1} \int T_{UU} dU\,.
\ee
The above result shows that $\Delta V$ becomes negative if $\int T_{UU} dU <0$. In this case, the shift $\Delta V$ corresponds to a time advance, and a signal coming from the left (right) asymptotic boundary can reach the right (left) asymptotic boundary of the geometry. In other words, if the ANEC is violated the wormhole becomes traversable.

GJW showed that ANEC can be violated if one introduces a double trace deformation that couples the two boundary theories \cite{Gao_2017}.
They computed the 1-loop expectation value of the bulk stress tensor in a BTZ black hole using a point splitting method. GJW's result can be written as follows \footnote{Here we simplify GJW's result a little bit by using the identity $\Gamma(\Delta+1/2)=\frac{\sqrt{\pi} \Gamma(2\Delta+1)}{2^{2\Delta}\Gamma(\Delta+1)}$.}
\be \label{eq-ANEC-GJW}
\int_{U_0}^{\infty}  T_{UU} dU = - \frac{h\, \Gamma(\Delta+\frac{1}{2})^2}{\pi (2\Delta+1)\Gamma(\Delta)^2} \frac{{}_2F_1\left(\frac{1}{2}+\Delta, \frac{1}{2}-\Delta ;\frac{3}{2}+\Delta;\frac{1}{1+U_0^2}\right)}{(1+U_0^2)^{\Delta+\frac{1}{2}}},
\ee
where the deformation is turned on at some time $t_0$, i.e., $h(t,{\bf x})=h \,\theta(t-t_0)$ and $U_0 = e^{t_0}$. Also, the traversability of the GJW wormhole can be diagnosed by a two-sided correlator that can be computed using the eikonal approximation~\cite{Maldacena_2017}.

In this section, we generalize the expression of ANE (\ref{eq-ANEC-GJW}) for a $(d+1)$ dimensional Rindler-AdS geometry using two different methods, namely, the point splitting method used in \cite{Gao_2017}, and the eikonal method used in \cite{Maldacena_2017}. We show that both methods give results that are consistent with each other. For future reference, we will use the following notation for the ANE
\begin{equation}
\mathcal{A}^{\infty} (U_0)\equiv \int_{U_0}^{\infty} T_{UU} dU\,,
\end{equation}
and change a subscript or superscript as occasion demands.

\subsection{Point splitting method} \label{sec-pointsplitting}
In this section, we compute the 1-loop expectation value of the stress tensor of the scalar field in the presence of the double trace deformation that couples the two asymptotic boundaries of the geometry. We consider the double-trace deformation, which corresponds to a time dependent piece in the Hamiltonian that is given by\footnote{The condition for $\delta H$ to be a relevant deformation is $\Delta_\mathcal{O} \leq d/2$, where $\Delta_\mathcal{O}$ is the scaling dimension of the operator $\mathcal{O}$. The unitarity bound implies $\Delta_\mathcal{O}>d/2-1$.}
\be
\delta H(t)= \int d {\bf x} \, h(t,{\bf x})\, \mathcal{O}_{L}(-t,{\bf x}) \mathcal{O}_R(t,{\bf x})\,,
\ee
where $d {\bf x} = \sinh^{d-2}\chi\, d\chi\, d\Omega_{d-2}$ and we take $h(t,{\bf x})= h \, \theta(t-t_0)$. This deformation induces a quantum correction in the matter stress tensor that can lead to a violation of the ANEC. 

For a scalar field with an action
\be
S_\text{scalar} = -\frac{1}{2}\int d^{d+1}x \sqrt{-g} \left(g^{\mu \nu} \partial_{\mu} \phi \partial_{\nu} \phi+m^2 \phi^2 \right),
\ee
the stress energy tensor can be obtained by varying the action with respect to $g^{\mu \nu}$
\be
T_{\mu \nu} = \partial_{\mu} \phi \partial_{\nu} \phi-\frac{1}{2} g_{\mu \nu} g^{\alpha \beta} \partial_{\alpha} \phi \partial_{\beta} \phi-\frac{1}{2}m^2 g_{\mu \nu}\phi^2\,.
\ee
The 1-loop expectation value of the stress energy tensor can be computed by point splitting 
\be \label{eq-splitting}
\langle T_{\mu \nu}\rangle = \lim_{x \rightarrow x'} \left( \partial_{\mu} \partial'_{\nu} G(x,x')-\frac{1}{2} g_{\mu \nu} g^{\alpha \beta} \partial_{\alpha} \partial'_{\beta} G(x,x')-\frac{1}{2}m^2 g_{\mu \nu}G(x,x')\right),
\ee
where $x$ and $x'$ denote bulk points, and $G(x,x')$ is a (renormalized) scalar two point function under the presence of the deformation
\bea 
G(x,x')&=& \langle \phi_R^{H}(t,r,{\bf x}) \phi_{R}^{H}(t',r',{\bf x'})\rangle \nonumber \\
&=& \langle U^{-1}(t,t_0)\phi_R^{I}(t,r,{\bf x}) U(t,t_0)U^{-1}(t',t_0) \phi_R^{I}(t',r',{\bf x'}) U(t',t_0)\rangle,
\eea
where $U(t,t_0)= \mathcal{T}\, e^{-i \int_{t_0}^{t} dt' \delta H(t')}$ denotes the evolution operator in the interaction picture. The subscript $R$ indicates a field in the right wedge, while the subscripts $H$ and $I$ indicate fields in the Heisenberg and interaction pictures, respectively.
By considering a small $h$ expansion, we write
\be
G(x,x') = G_0(x,x')+G_1(x,x')\, h
+\mathcal{O}(h^2)\,,\\
\ee
and we use (\ref{eq-splitting}) to compute $\langle T_{\mu \nu}\rangle$ as
\be
\langle T_{\mu \nu}\rangle= \langle T_{\mu \nu}\rangle_0+\langle T_{\mu \nu}\rangle_1 \,h+\mathcal{O}(h^2)\,.
\ee

In the Rindler $AdS_{d+1}$ background, the 1-loop contribution to the bulk two point function, evaluating at $V=0$ in Kruskal coordinates, is given by
\begin{align}\label{G1UU}
G_1(U,U',\mathbf{x},\mathbf{x'}) =& \,\, 2 \sin{(\pi \Delta)} c_{\Delta}^2 \int\frac{dU_1}{U_1} \int d \mathbf{x}_1 h(U_1,\mathbf{x}_1)\, \theta\bigg(\frac{U}{U_1}-\cosh{d(\mathbf{x},\mathbf{x}_1)} \bigg) \\
&  \times \bigg( \frac{1}{\frac{U}{U_1}-\cosh{d(\mathbf{x},\mathbf{x}_1)}}\bigg)^\Delta \bigg( \frac{1}{U' U_1 +\cosh{d(\mathbf{x'},\mathbf{x}_1)}}\bigg)^\Delta 
     + (U,\mathbf{x} \leftrightarrow U',\mathbf{x}')   \nonumber \\
     \equiv& \,\,F(U,U',\mathbf{x},\mathbf{x'})+F(U',U,\mathbf{x},\mathbf{x'}) \,,\nonumber
\end{align}
where $d(\mathbf{x},\mathbf{x_1})$ is the geodesic distance between $\mathbf{x}$ and $\mathbf{x}_1$ in $\mathbb{H}_{d-1}$.



\begin{figure}
\begin{center}
\begin{tabular}{cc}
\setlength{\unitlength}{1cm}
\hspace{-0.9cm}
\includegraphics[width=7.5cm]{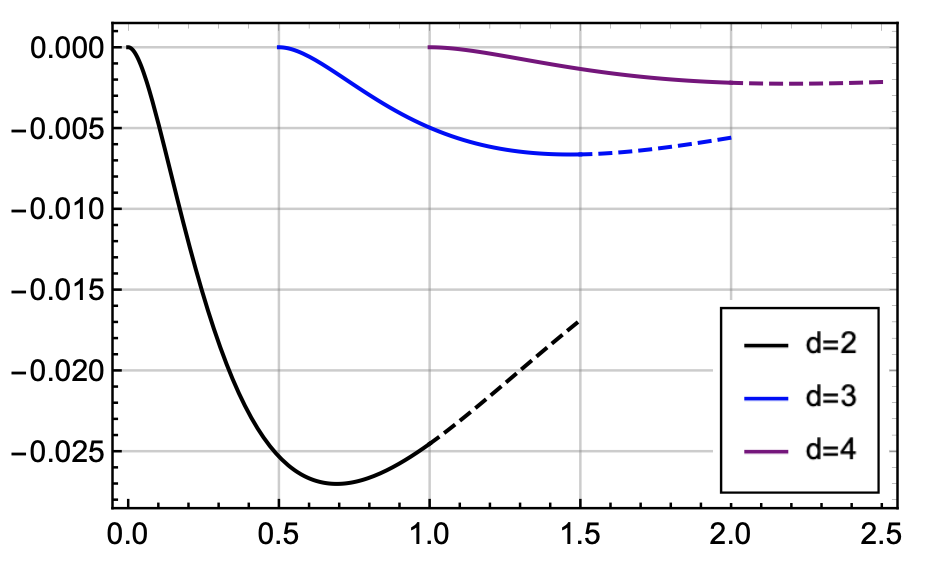}
\qquad\qquad &
\includegraphics[width=7.5cm]{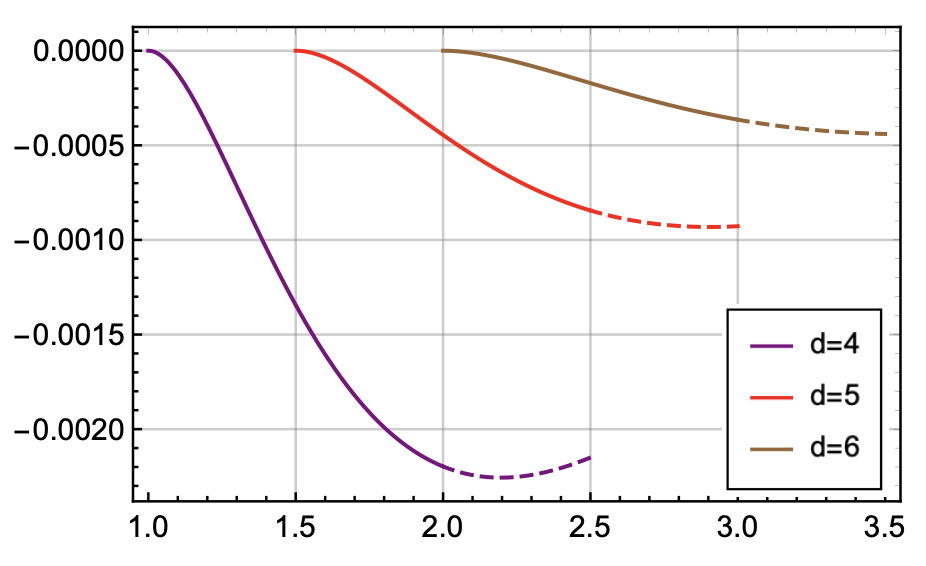}
\qquad
\put(-100,-10){$\Delta$}
\put(-355,-10){$\Delta$}
\put(-480,50){\rotatebox{90}{$\mathcal{A}^{\infty} (1)$}}
\put(-228,50){\rotatebox{90}{$\mathcal{A}^{\infty} (1)$}}
\end{tabular}
\end{center}
\caption{$\mathcal{A}^{\infty} (U_0) = \int_{U_0}^{\infty}T_{UU} dU$ versus $\Delta$ for increasing values of $d$. On the left panel we have $d=$ 2 (black curve), $d=3$ (blue curve), $d=4$ (purple curve). On the right panel we have $d=4$ (purple curve), $d=5$ (red curve), $d=6$ (brown curve). The solid curves are plotted for $\frac{d}{2}-1 \leq \Delta \leq \frac{d}{2}$, which is the range in which the deformation is relevant, while the dashed ones are plotted for $\frac{d}{2}-1 \leq \Delta \leq \frac{d+1}{2}$, which is the range in which (\ref{ANEC_dUTUU}) is valid. Here, we set $U_0=1$.}
    \label{fig:ANEC}
\end{figure}

For simplicity, we set $\mathbf{x}=\mathbf{x'}=0$ so that the geodesic distance is given by $d({\bf x_1,x})=\chi_1$. This is equivalent to consider homogeneous perturbations, in which case $G_1$ does not depend on ${\bf x}$ and ${\bf x'}$ and we can conveniently set them to zero.
By defining $y=\cosh{\chi_1}$, we obtain the $UU$-component of the stress energy tensor on the horizon as follows
\begin{align} \label{eq-stress}
T_{UU}=&2 \lim_{U'\rightarrow U} \partial_{U} \partial_{U'} F(U,U')\\
 =& -2\Delta c_{\mathcal{O}}\, h \,\text{vol}(S_{d-2}) \lim_{U'\rightarrow U} \partial_U \int_{U_0}^U \frac{dU_1}{U_1} \int_{1}^{U/U_1} \frac{dy}{ (y^2-1)^{\frac{3-d}{2}}}\frac{U_1^\Delta}{(U-U_1 y)^\Delta (U' U_1+y)^{\Delta+1}} \,, \nonumber
\end{align}
where $c_\mathcal{O}= 2 \sin (\pi \Delta) c_{\Delta}^2$. With the above expression, we show in Appendix \ref{app:A} that
\begin{align}\label{ANEC_dUTUU}
\mathcal{A}^{\infty} (U_0)
= -{\text{vol}(S_{d-2})} \frac{h \pi^{\frac{1-2d}{2}} \Gamma(\frac{d-1}{2} )}{2(2\Delta +1)}\frac{\Gamma(\Delta+\frac{1}{2}) \Gamma(\Delta+ \frac{3-d}{2})}{\Gamma(\Delta+1-\frac{d}{2} )^2}
 \frac{{}_2F_1\bigg(\Delta+\half, \half-\Delta, \Delta+\frac{3}{2};\frac{1}{1+U_0^2}\bigg)
}{ (1+U_0^2)^{\Delta+\frac{1}{2}}}  \,.
\end{align}
Note that we recover GJW result \eqref{eq-ANEC-GJW} by setting $d=2$ in \eqref{ANEC_dUTUU}. 
The unitarity bound implies $\Delta \geq \frac{d}{2}-1$, while the condition for the deformation to be relevant reads $\Delta <\frac{d}{2}$. The derivation of (\ref{ANEC_dUTUU}) using the point splitting method requires $\Delta < \frac{d+1}{2}$ (see Appendix \ref{app:A}), but the same formula can be obtained using the eikonal method (see Sec.~\ref{sec-eikonal}) without any upper bound on $\Delta$. Fig.~\ref{fig:ANEC} shows $\mathcal{A}^{\infty} (U_0)$ versus $\Delta$ for increasing values of $d$. The violation of ANEC quickly decreases as we increase the dimensionality of the spacetime. This suggests that it is more difficult to send information through the wormhole in higher dimensional cases, as compared to lower dimensional cases. We will confirm that this is indeed the case in Sec.~\ref{sec-bound}, where we study bounds on information transfer. 

Note that $\mathcal{A}^{\infty}(U_0)$ is related to a perturbation in which $h(t,{\bf x})=h \,\theta(t- t_0)$. It is also convenient to consider an instantaneous perturbation $h^\text{inst}(t,{\bf x}) = h\, \delta(t-t_0)$.
In this case, the average null energy is
\begin{equation}
\mathcal{A}^\text{inst}=-U_0 \partial_{U_0}	 \mathcal{A}^{\infty} (U_0).
\end{equation}
By direct differentiation of \eqref{ANEC_dUTUU}, we obtain
\begin{equation}\label{ANEC_inst}
	\mathcal{A}^\text{inst}(U_0) = - h\, \text{vol}(S_{d-2})\, \frac{\pi^{\frac{1-2d}{2}}}{2}\frac{\Gamma(\frac{d-1}{2}) \Gamma(\Delta+\frac{1}{2}) \Gamma(\Delta +\frac{3-d}{2})}{\Gamma(\Delta +1-\frac{d}{2})^2} \bigg( \frac{U_0}{1+U_0^2}\bigg)^{2\Delta+1} \,.
\end{equation}
This provides a higher-dimensional generalization ($d \geq 3$) of the results for a BTZ black hole derived in \cite{Freivogel_2020}
\begin{equation}\label{ANEC_inst_BTZ}
\mathcal{A}_{d=2}^\text{inst}(U_0) = - \frac{h \Gamma(\Delta+\frac{1}{2})^2}{\pi \Gamma(\Delta)^2} \bigg( \frac{U_0}{1+U_0^2}\bigg)^{2\Delta+1}.
\end{equation}

In the next section, we show that \eqref{ANEC_dUTUU} and \eqref{ANEC_inst} can also be obtained by using the eikonal approximation, as done in \cite{Maldacena_2017} for a two-dimensional gravitational system.

\subsection{Eikonal method} \label{sec-eikonal}

In this section, we analyze the traversability wormholes in higher dimensions ($d \geq 2$) by following the approach of \cite{Maldacena_2017}, which uses the eikonal approximation to directly compute the expectation value of a two-sided correlation function of the form
\be \label{eq-V}
\langle \left[ \psi_{L} (-t_1,{\bf x_1}),  e^{-i  \mathcal{V}} \psi_{R}(t_2,{\bf x_2}) e^{i  \mathcal{V}} \right] \rangle,
\ee
where the expectation value is taken in a thermofield double state, and
\be \label{eq-doubletrace}
\mathcal{V}=\frac{1}{K} \sum^{K}_{i=1}\int dt' \, d{\bf x'}h(t',\textbf{x}') \mathcal{O}_{L}^{i}(-t',\textbf{x}') \mathcal{O}_{R}^{i}(t',\textbf{x}') 
\ee
is a double trace deformation involving $K$ light fields. It is convenient to consider more than one field because the large $K$ limit leads to simplifications.

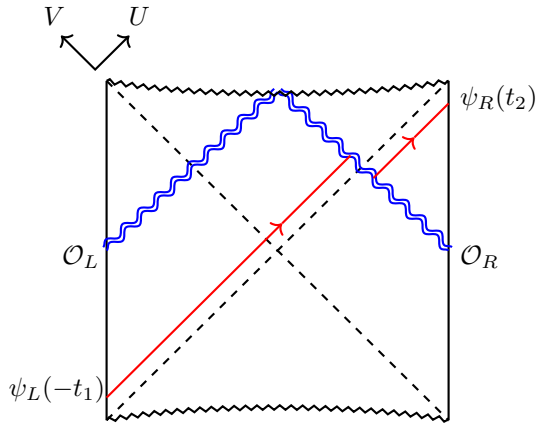
\begin{figure}
\centering

\begin{tikzpicture}[scale=1.5]
\draw [thick]  (0,0) -- (0,3);
\draw [thick]  (3,0) -- (3,3);
\draw [thick,dashed]  (0,0) -- (3,3);
\draw [thick,dashed]  (0,3) -- (3,0);

\draw [thick,blue,decorate,decoration={snake,segment length=3mm,amplitude=0.5mm}]  (3,1.5) -- (1.5,2.9);
\draw [thick,blue,decorate,decoration={snake,segment length=3mm,amplitude=0.5mm}]  (3.03,1.53) -- (1.53,2.93);

\draw [thick,blue,decorate,decoration={snake,segment length=3mm,amplitude=0.5mm}]  (0,1.5) -- (1.5,2.9);
\draw [thick,blue,decorate,decoration={snake,segment length=3mm,amplitude=0.5mm}]  (-0.03,1.53) -- (1.47,2.93);

\draw [red,thick,->]  (0,0.2) -- (1.54,1.74);
\draw [red,thick]  (1.54,1.74) -- (2.14,2.34);
\draw [red,thick,->]  (2.34,2.14) -- (2.7,2.5);
\draw [red,thick]  (2.7,2.5) -- (3,2.8);

\draw [thick,decorate,decoration={zigzag,segment length=1.5mm, amplitude=0.3mm}] (0,3) .. controls (.75,2.85) 
and (2.25,2.85) .. (3,3);
\draw [thick,decorate,decoration={zigzag,segment length=1.5mm,amplitude=.3mm}]  (0,0) .. controls (.75,.15) and (2.25,.15) .. (3,0);

\draw[thick,<->] (-0.4,3.4) -- (-0.1,3.1) -- (0.2,3.4);

\end{tikzpicture}
\vspace{0.1cm}
\put(-150,60){\rotatebox{0}{\small $\mathcal{O}_L$}}
\put(-1,60){\rotatebox{0}{\small $\mathcal{O}_R$}}
\put(-170,10){\rotatebox{0}{\small $\psi_L(-t_1)$}}
\put(-1,120){\rotatebox{0}{\small $\psi_R(t_2)$}}
\put(-157,150){\rotatebox{0}{\small $V$}}
\put(-125,150){\rotatebox{0}{\small $U$}}

\caption{\small The non-local coupling between $\mathcal{O}_L$ and $\mathcal{O}_R$ introduces a negative-energy shock wave in the bulk that makes the wormhole traversable. The traversability can be diagnosed by a two-sided correlation function (\ref{eq-V}) involving $\psi_L$ and $\psi_R$.}
\label{fig-info-tranfer}
\end{figure}

The correlation function \eqref{eq-V} measures the response of $\psi_R$ to a perturbation on the left side of the geometry. It takes a non-zero value when the signal can travel through the wormhole and reach the right boundary \cite{Maldacena_2017}. The double-trace deformation $\mathcal{V}$ introduces a {\it negative-energy shock wave} in the bulk, which makes geodesics crossing the shock wave suffer a negative shift in the $V$ direction\footnote{This should be contrasted with the positive shift introduced by a positive-energy shock wave. }. This negative shift is responsible for making the bulk field corresponding to the operator $\psi_L$ correlate with the operator $\psi_R$, as shown in Figure \ref{fig-info-tranfer}. As we will see in the following, \eqref{eq-V} contains information about the ANEC violation and provides a method to compute ANE, which is perfectly consistent with the one by the point-splitting method in \cite{Gao_2017}.

It turns out to be more convenient to work with the following correlation function
\be \label{eq-C}
C=\langle e^{-i\mathcal{V}}\psi_R(t_2,{\bf x_2})e^{i\mathcal{V}}\psi_L(-t_1,{\bf x_1})\rangle,
\ee
whose imaginary part gives the original commutator $\langle \left[ \psi_{R}, e^{-i \mathcal{V}} \psi_{L} e^{i  \mathcal{V}} \right] \rangle = -2 \,i\, \text{Im} (C)$.
In the large $K$ and small $G_N$ limits, $C$ takes a simple form~\cite{Maldacena_2017}
\be \label{eq-CtildefromC}
C=e^{-i\langle \mathcal{V} \rangle}\tilde{C} \,, \, \, \, \tilde{C} \equiv \langle \psi_{R}(t_2,{\bf x_2}) e^{i \mathcal{V}} \psi_{L}(-t_1,{\bf x_1}) \rangle\,.
\ee
The correlator $\tilde{C}$ has all the information we need about the traversability of the wormhole. For simplicity, let us first consider a small $h$ (recall that $\mathcal{V} \sim h$) expansion and compute the result at linear order in $h$
\be \label{eq-Ctildefirstorder}
\tilde{C}_{1}=\frac{i }{K} \sum_{j=1}^{K} \int_{0}^{t} d t' \int d {\bf x'} \langle \psi_{R} (t_{2}, {\bf x_2}) \mathcal{O}_{L}^{j} (-t', {\bf x'}) \mathcal{O}_{R}^{j} (t',{\bf x'}) \psi_L (-t_1,{\bf x_1}) \rangle \,h(t',{\bf x'}),
\ee
which is basically an out-of-time-order correlator that can be computed using the techniques introduced in \cite{Shenker:2014cwa}. For convenience, we omit the sum in $K$ for now and consider $K=1$. Explicit expressions involving $K$ will be reintroduced when the large $K$ approximation is needed.

 By following \cite{Shenker:2014cwa, almheiri2018escaping}, we now review and generalize the method of \cite{Maldacena_2017} for cases in which $d \geq 2$. 
We first write $\tilde{C}_{1}$ as an amplitude
\be
\tilde{C}_{1} = i \,\langle \text{out} | \text{in} \rangle,
\ee
where the `in' and `out' states are defined as follows
\be
| \text{in} \rangle = \mathcal{O}_{R}(t',{\bf x'}) \psi_{L} (-t_1, {\bf x_1}) | \beta \rangle, \,~~\,\,
| \text{out} \rangle = \mathcal{O}_{L}^{\dagger}(-t',{\bf x'}) \psi_{R}^{\dagger} (t_2, {\bf x_2}) | \beta \rangle .
\ee
Then, we expand $| \text{in} \rangle$ and $| \text{out} \rangle$ in a basis of well-defined momentum in the $U$ or $V$ direction and well-defined transverse position ${\bf \tilde{x}} \in \mathbb{H}_{d-1}$
\bea
&&| \text{in} \rangle = \int dp_{1}^{U} dp_{4}^{V} \int d{\bf \tilde{x}_{1}} d{\bf \tilde{x}_{4}}  \Psi_{\mathcal{O}_{R}}(p_{4}^{V},{\bf \tilde{x}_{4}}) \Psi_{\psi_{L}}(p_{1}^{U},{\bf \tilde{x}_{1}}) \left( | p_{4}^{V},{\bf \tilde{x}_{4} }\rangle \otimes | p_{1}^{U},{\bf \tilde{x}_{1}} \rangle \right)_{\text{in}},\\
&&|\text{out} \rangle = \int  dp_{2}^{U} dp_{3}^{V} \int d{\bf \tilde{x}_{2}} d{\bf \tilde{x}_{3}}  \Psi_{\mathcal{O}_{L}}(p_{3}^{V},{\bf \tilde{x}_{3}})  \Psi_{\mathcal{\psi}_{R}}(p_{2}^{U},{\bf \tilde{x}_{2}}) \left( | p_{3}^{V},{\bf \tilde{x}_{3}} \rangle \otimes | p_{2}^{U}, {\bf \tilde{x}_{2}} \rangle \right)_{\text{out}},
\eea
where the integral is over all the exposed variables. The wave functions are given by Fourier transforms of bulk-boundary propagators along either the $U=0$ or $V=0$ horizons
\begin{align}
\begin{split}
&\Psi_{\mathcal{O}_{R}}(p_{4}^{V},{\bf \tilde{x}_4})=\int dU e^{ia_{0} p^{V}_{4} U/2}\langle \phi_{\mathcal{O}}(U,V, {\bf \tilde{x}_4}) \mathcal{O}_{R} (t',{\bf x'})\rangle_{V=0},\\
&\Psi_{\psi_{L}}(p_{1}^{U},{\bf \tilde{x}_1})=\int dV e^{ia_{0} p^{U}_{1} V/2}\langle \phi_{\psi}(U,V,{\bf \tilde{x}_1}) \psi_{L} (-t_1,{\bf x_1})\rangle_{U=0},\\
&\Psi_{\psi_{R}}(p_{2}^{U},{\bf \tilde{x}_2})=\int dV e^{ia_{0} p^{U}_{2} V/2}\langle \phi_{\psi}(U,V,{\bf \tilde{x}_2}) \psi_{R}^{\dagger} (t_2,{\bf x_2})\rangle_{U=0},\\
&\Psi_{\mathcal{O}_{L}}(p_{3}^{V},{\bf \tilde{x}_3})=\int dU e^{ia_{0} p^{V}_{3} U/2}\langle \phi_{\mathcal{O}}(U,V, {\bf \tilde{x}_3}) \mathcal{O}_{L}^{\dagger} (-t', {\bf x'})\rangle_{V=0},
\end{split}
\end{align}
where $a_0$ denotes the $U V$ component of the metric either at $U=0$ or at $V=0$ \footnote{For a Rindler-AdS geometry of the form (\ref{eq-metric-kruskal}), $a_0=4$. For convenience, we keep the parameter $a_0$ in our expressions.}.
Here, $\phi_\mathcal{O}$ and $\phi_\psi$ denote the bulk fields which are dual to the boundary operators $\mathcal{O}$ and $\psi$, respectively. The basis is normalized as 
\be
\langle p^{U},{\bf x}|q^{U},{\bf x'}\rangle = \frac{a_{0}^{2}\, p^{U}}{4 \pi r_{0}^{d-1}} \delta(p^{U}-q^{U})\delta({\bf x,x'}),
\ee
where $\delta({\bf x,x'})$ denotes a delta function in $(d-1)$ dimensional hyperbolic space
\be
\delta({\bf x,x'})=\frac{\delta(\chi-\chi')}{\sinh^{d-2}{\chi}} \frac{\delta(\theta_{1}-\theta_{1}') \cdots \delta(\theta_{d-3} -\theta_{d-3}')}{\sin{\theta_{1}} \cdots \sin^{d-3}{\theta_{d-3}}} \delta(\phi-\phi').
\ee
By using the bulk-boundary propagator (\ref{eq-bbprop}), which is valid for both $(U,V,\bf x)$ and $(t',{\bf x'})$ in the right exterior region, we obtain the following wave functions
\begin{align} \label{eq-wave function}
\begin{split}
&\Psi_{\mathcal{O}_{R}}(p_{4}^{V}, {\bf \tilde{x}_{4}})=\Theta(p_{4}^{V})\frac{2 \pi i c_{\mathcal{O}} \, e^{t'}}{\Gamma(\Delta_{\mathcal{O}})} \bigg(\frac{-i a_{0} p_{4}^{V} e^{t'}}{2}\bigg)^{\Delta_{\mathcal{O}}-1}e^{i\frac{a_{0}}{2}p_{4}^{V} e^{t'} \cosh{d({\bf \tilde{x}_{4},x'})}},\\
&\Psi_{\psi_{L}}(p_{1}^{U}, {\bf \tilde{x}_{1}})=\Theta(p_{1}^{U})\frac{2 \pi i c_{\psi}  \,e^{t_1}}{\Gamma(\Delta_{\psi})} \bigg(\frac{-i a_{0} p_{1}^{U} e^{t_1}}{2} \bigg)^{\Delta_{\psi}-1} e^{i \frac{a_{0}}{2} p_{1}^{U} e^{t_1} \cosh{d({\bf \tilde{x}_{1},x_1})}},\\
&\Psi_{\psi_{R}}(p_{2}^{U}, {\bf \tilde{x}_2})=\Theta(p_{2}^{U})\frac{2\pi i c_{\psi} \,e^{-t_2^{*}}}{\Gamma(\Delta_{\psi})} \bigg(\frac{-i a_{0} p_{2}^{U} e^{-t_2^{*}}}{2}\bigg)^{\Delta_{\psi}-1} e^{i\frac{a_{0}}{2} p_{2}^{U} e^{-t_2^{*}}\cosh{d({\bf \tilde{x}_2,x_2})}},\\
&\Psi_{\mathcal{O}_{L}}(p_{3}^{V}, {\bf \tilde{x}_3})=\Theta(p_{3}^{V})\frac{2\pi i c_{\mathcal{O}}e^{-t'^*}}{\Gamma(\Delta_{\mathcal{O}})} \bigg(\frac{-i a_{0} p_{3}^{V} \,e^{-t'^*}}{2}\bigg)^{\Delta_{\mathcal{O}}-1}e^{i\frac{a_{0}}{2}p_{3}^{V} e^{-t^{*}}\cosh{d({\bf \tilde{x}_3,x'})}},
\end{split}
\end{align}
where we used the Hankel representation of the Gamma function $\frac{1}{\Gamma(z)}=\frac{1}{2 \pi i} \int \tau^{-z}e^{\tau} d\tau$.
Now we use the eikonal approximation to write
\begin{align}
_\text{out}\left( \langle p_{3}^{V},{\bf \tilde{x}_{3}}  | \langle p_{2}^{U}, {\bf \tilde{x}_{2} }|\right) & \left( | p_{4}^{V},{\bf \tilde{x}_{4}} \rangle | p_{1}^{U}, {\bf \tilde{x}_{1}} \rangle \right)_\text{in} \nonumber\\
&=\left(\frac{a_{0}^{2}}{4 \pi} \right)^{2} p_{1}^{U} p_{4}^{V} e^{i \delta} \delta ( p_{1}^{U}-p_{2}^{U} ) \delta ( p_{3}^{V}-p_{4}^{V} ) \delta ({\bf \tilde{x}_{1},\tilde{x}_{2}} ) \delta ({\bf \tilde{x}_{3},\tilde{x}_{4}} )\,,    
\end{align}
where the phase shift $\delta$ is given by the sum of the classical actions of the fields $\phi_{\psi}$ and $\phi_{\mathcal{O}}$. In our setup, we obtain
\be \label{eq-phaseshiftlocal}
\delta = 4 \pi G_{N}a_{0}\, p_{1}^{U} p_{4}^{V}\, f({\bf \tilde{x}_1,\tilde{x}_4 })\,.
\ee
We will review the calculation of transverse profile $f({\bf \tilde{x}_1,\tilde{x}_4 })$ in Sec.~\ref{sec-bound}.
Then, at first order in $h$ we can write $\tilde{C}_{1}$  as
\bea
\tilde{C}_{1}&=& \alpha^{2} \int d p_{1}^{U} d {\bf \tilde{x}_{1}} \left[ p_{1}^{U} \Psi_{\psi_{R}}^{*} (p_{1}^{U}, {\bf \tilde{x}_{1}}) \Psi_{\psi_{L}}(p_{1}^{U},{\bf \tilde{x}_{1}})\right]\nonumber\\
&&\times \int d p_{4}^{V} d {\bf \tilde{x}_{4}} \int d t' d {\bf x'} \left[h (t',{\bf x'}) p_{4}^{V} \Psi^*_{\mathcal{O}_{L}}(p_{4}^{V},{\bf \tilde{x}_{4}}) \Psi_{\mathcal{O}_{R}}(p_{4}^{V},{\bf \tilde{x}_{4}}) e^{i\delta} \right],
\eea
where $d {\bf x} = d \chi \sinh^{d-2}{\chi} d\Omega_{d-2}$ and $\alpha = \frac{a_{0}^2}{4 \pi}$. 

At all orders in $h$, one can show that the result exponentiates~\cite{Maldacena_2017}
\be \label{eq-correlator}
\tilde{C}=\alpha \int d q\, d {\bf \tilde{x}_{1}} q\, \Psi_{\psi_{R}}^{*}(q,{\bf \tilde{x}_{1}}) \Psi_{\psi_{L}} (q,{\bf \tilde{x}_{1}})\, e^{-i D},
\ee
where
\be \label{eq-expD}
D= - \alpha \int d p d {\bf \tilde{x}_{4}} \int d t' d {\bf x'} h(t',{\bf x'}) p\, \Psi_{\mathcal{O}_{L}}^{*} (p,{\bf \tilde{x}_{4}}) \Psi_{\mathcal{O}_{R}} (p,{\bf \tilde{x}_{4}})  e^{i \delta},
\ee
where we replace $p_{1}^{U}$ and $p_{4}^{V}$ by $q$ and $p$, respectively.
Using the explicit form of the wave functions (\ref{eq-wave function}), we find
 \be \label{eq-Ctilde}
 \tilde{C}=-\alpha \frac{2^{2\Delta_{\psi}} \pi^2 c_{\psi}^2 }{\Gamma({\Delta_{\psi}})^2}\int d q \, d {\bf \tilde{x}_{1}} q^{2\Delta_{\psi}-1} e^{2 i q \left( e^{-t_{2}} \cosh{d({\bf \tilde{x}_{1},x_{2}})} + e^{t_{1}} \cosh{d({\bf \tilde{x}_{1},x_{1}})} \right) } e^{-i \pi \Delta_{\psi}} e^{- i D},
 \ee
 with
 \be \label{eq-expD2}
 D= \alpha  \frac{\pi^{2} c_{\mathcal{O}}^{2} 2^{2\Delta_{\mathcal{O}}}}{\Gamma(\Delta_{\mathcal{O}})^{2}} \int d p \, d {\bf \tilde{x}_{4}} \int d t'\, d {\bf x'} p^{2\Delta_{\mathcal{O}}-1} e^{4 i p \cosh{d({\bf \tilde{x}_{4},x'})} \cosh{t'}} e^{-i \pi \Delta_{\mathcal{O}}} e^{i \delta}.
 \ee
Let us first compute the exponent $D$. Evaluating the integral with respect to $p$, we find
\be
D = \alpha\,  \,2^{4 \Delta_{\mathcal{O}}} b_{\mathcal{O}}^{2}\Gamma(2 \Delta_{\mathcal{O}}) \int  d {\bf \tilde{x}_{4}} \int d t' d {\bf x'} \frac{ h(t',{\bf x'})}{\big[ 4 \cosh{d({\bf \tilde{x}_{4},x'}) \cosh{t'} + 16 \pi G_N q f({\bf \tilde{x}_1,\tilde{x}_4 })} \big]^{2 \Delta_{\mathcal{O}}}},
\ee
where $b_{\mathcal{O}}=\frac{\pi c_{\mathcal{O}}}{2^{\Delta_{\mathcal{O}}}\Gamma(\Delta_{\mathcal{O}})}$. 
\paragraph{Probe limit}
Here, we consider the probe limit, in which the backreaction of the signal is too small to deform the negative-energy shock wave geometry. We implement this approximation by performing an expansion to the first order in $G_{N} q$ and evaluating $D$ order by order.
We find $D=D_{0}+D_{1} q$, where
\be
D_{0}=\alpha \,b_{\mathcal{O}}^2 \Gamma(2\Delta_{\mathcal{O}}) \int d t' d {\bf x'} d{\bf \tilde{x}_{4}} \frac{ h(t',{\bf x'})}{ \big[\cosh{d({\bf \tilde{x}_{4},x'})} \cosh{t'} \big]^{2\Delta_{\mathcal{O}}}}\,,
\ee
and
\be \label{eq-D1}
D_{1} = -\alpha \, b_{\mathcal{O}}^{2} \frac{\Delta_{\mathcal{O}} \Gamma(2\Delta_{\mathcal{O}})}{2} \int dt' d{\bf x'} d {\bf \tilde{x}_4}\frac{ 16 \pi G_N f({\bf \tilde{x}_1,\tilde{x}_4 })\, h(t',{\bf x'})}{\big[ \cosh{t'} \cosh{d({\bf \tilde{x}_{4},x'})} \big]^{2 \Delta_{\mathcal{O}} + 1}}\,.
\ee
The correlator in the probe approximation can then be written as
\begin{align} \label{eq-DeltaV}
C_\text{probe} &\equiv e^{-i  \langle \mathcal{V} \rangle} \tilde{C} \,|_{D=D_0+q D_1} \nonumber\\
&=\langle \psi_{R} e^{-i D_{1} q} \psi_{L} \rangle=\langle \psi_{R} e^{i D_{1} \frac{2}{a_0} \hat{p}_{V}} \psi_{L} \rangle = \langle \psi_{R} e^{i \Delta V \hat{p}_{V}} \psi_{L} \rangle \,.
\end{align}
In the first line, the zeroth order term $D_{0}$ corresponds to $-\langle \mathcal{V} \rangle$ and cancels the overall factor of $e^{-i  \langle \mathcal{V} \rangle}$ in the correlator. In the second line we used that the momentum of the signal along the horizon is $q$.
The above result shows that $D_{1}$ corresponds to a shift in the $V$ direction, i.e. $\Delta V=\frac{2}{a_0} D_{1}$. The ANE can then be computed as
\be \label{eq-ANED1}
\int T_{UU} dU =\frac{ d-1}{4 \pi G_{N}}  \Delta V = \frac{d-1}{8 \pi G_{N}} D_{1},
\ee
where we used \eqref{eq-deltav-ANEC} and $a_0=4$, which is appropriate for a Rindler AdS geometry. 

Note that the negative-energy shock wave renders the wormhole traversable. The backreaction of the bulk fields corresponding to the operators $\psi_L$ and $\psi_R$ can be described by a positive-energy shock wave geometry which have the tendency to close the wormhole.

Using (\ref{eq-ANED1}) and (\ref{eq-D1}), we can evaluate the ANE and compare the result with the one obtained by the point-splitting method in Sec.~\ref{sec-pointsplitting}. The consistency between the two different methods was checked numerically for a rotating BTZ black hole in \cite{Caceres_2018}. In the following, we will show the equivalence between both methods by finding an explicit analytic formula for the ANE using the eikonal approximation.

 \subsubsection{Homogeneous perturbations} \label{sec-homogeneous}
In this section, we consider the case in which the double trace deformation produces a shock wave that is homogeneous in the transverse space, i.e, the shock wave transverse profile does not depend on the coordinates ${\bf x} \in \mathbb{H}_{d-1}$. The signal, on the other hand, we consider to be produced by a local operator. In this case, we can derive a formula for $D_1$ that is very similar to $(\ref{eq-D1})$.

First, we expand the initial and final states of the field excitations produced by the operators $\mathcal{O}_R$ and $\mathcal{O}_L$ in a basis of well-defined momentum $| p \rangle$, instead of $|p,{\bf \tilde{x}} \rangle$. Second, we remove the ${\bf \tilde{x}}$ dependence of the phase shift, which we take as follows\footnote{For homogeneous shocks, the shock wave transverse profile does not depend on the coordinates in the transverse space, and $f({\bf \tilde{x}_1,\tilde{x}_4 })$ is replaced by $\frac{1}{d-1}$. See Sec.~\ref{sec-bound}.} $\delta_\text{hom}=4 \pi G_N a_0  p_1^{V} p_4^{U}/(d-1)$.  Proceeding as before, we can show that
  \be \label{eq-D1hom}
D_{1} = -\alpha \, b_{\mathcal{O}}^{2} \frac{\Delta_{\mathcal{O}} \Gamma(2\Delta_{\mathcal{O}})}{2 (d-1)} \int dt' d{\bf x'} \frac{ 16 \pi G_N h(t',{\bf x'})}{\big[ \cosh{t'} \cosh{d(\bf{x'},0)} \big]^{2 \Delta_{\mathcal{O}} + 1}}\,.
\ee
To simplify the calculation of $D_{1}$, we ignore the dependence on the coordinates on the sphere $S^{d-2}$, and write the geodesic distances as $d({\bf x,x'})=\chi-\chi'$. Moreover, we consider an instantaneous perturbation $h(t',{\bf x'})=h\,\delta(t'-t_{0})$. Then, by direct integration of (\ref{eq-D1hom}) and using (\ref{eq-ANED1}), we obtain\footnote{Here we used the identities $\int_{0}^{\infty} d \chi \frac{\sinh^{d-1} \chi}{(\cosh \chi)^{2 \Delta+1}}=\frac{1}{2}\frac{\Gamma\left(\frac{d-1}{2}\right) \Gamma \left( \Delta+\frac{3-d}{2}\right)}{\Gamma(\Delta+1)}$ and $\Gamma(2\Delta)=\frac{\Gamma(\Delta)\Gamma(\Delta+1/2)}{\sqrt{\pi} 2^{1-2\Delta}}$.}
\be \label{eq-ANE_inst}
\mathcal{A}^\text{inst}(U_0)=- h \text{vol}(S_{d-2}) \frac{\pi^{\frac{1-2d}{2}}}{2}\frac{\Gamma(\frac{d-1}{2})\Gamma(\Delta_{\mathcal{O}}+\frac{1}{2})\Gamma(\Delta_{\mathcal{O}}+\frac{3-d}{2})}{\Gamma(\Delta_{\mathcal{O}}-\frac{d-2}{2})^{2}} \bigg( \frac{U_0}{1+ U_0^2} \bigg)^{2\Delta_{\mathcal{O}} +1} \,,
\ee
 where $U_0=e^{t_0}$. The above result perfectly matches the result obtained by point splitting for an instantaneous perturbation (\ref{ANEC_inst}). 

We can now consider the case where $h(t',{\bf x'})=h\, \theta(t'-t_{0})$. We just have to use the relation

\be \label{eq-ANErel}
\mathcal{A}^{\infty}(U_{0})=\int_{U_{0}}^{\infty} \frac{dU}{U} \mathcal{A}^{\text{inst}} (U)
\ee
By (\ref{eq-ANErel}) and using the identity $\Gamma(\Delta+1-\frac{d}{2}) = (\Delta-\frac{d}{2})\Gamma(\Delta-\frac{d}{2})$, we can show that
\bea \label{eq-ANEC step ftn}
\mathcal{A}^{\infty}(U_{0})&= -h \, &\text{vol}(S_{d-2}) \frac{\pi^{\frac{1-2d}{2}}}{2}\\ && \times \frac{\Gamma\left(\frac{d-1}{2}\right)\Gamma\left(\Delta+\frac{1}{2}\right)\Gamma\left(\Delta+\frac{3-d}{2}\right)}{\Gamma\left(\Delta+1-\frac{d}{2}\right)^{2}}\frac{_{2}F_{1}\bigg(\frac{1}{2}+\Delta,\frac{1}{2}-\Delta,\frac{3}{2}+\Delta,\frac{1}{1+U_{0}^{2}}\bigg)}{(2\Delta+1)(1+U_{0}^{2})^{\Delta+\frac{1}{2}}}.\nonumber
\eea
 The above result perfectly matches the formula for the ANE obtained via point-splitting in (\ref{ANEC_dUTUU}), and it provides a generalization of the results of \cite{Gao_2017} and \cite{Maldacena_2017} to higher dimensions ($d \geq 2$).

We recall that $D_1=\Delta V$ characterizes the null shift that a probe particle undergoes when crossing the shock wave produced by the double trace deformation. When $\Delta V <0$, a signal can be transmitted through the wormhole, producing non-trivial correlations between left and right boundary operators, which can be measured by correlators of the form (\ref{eq-C}).

\begin{figure}
\begin{center}
    \subfigure[$t_1=1.001T_c $] 
    {\includegraphics[width=7.1cm]{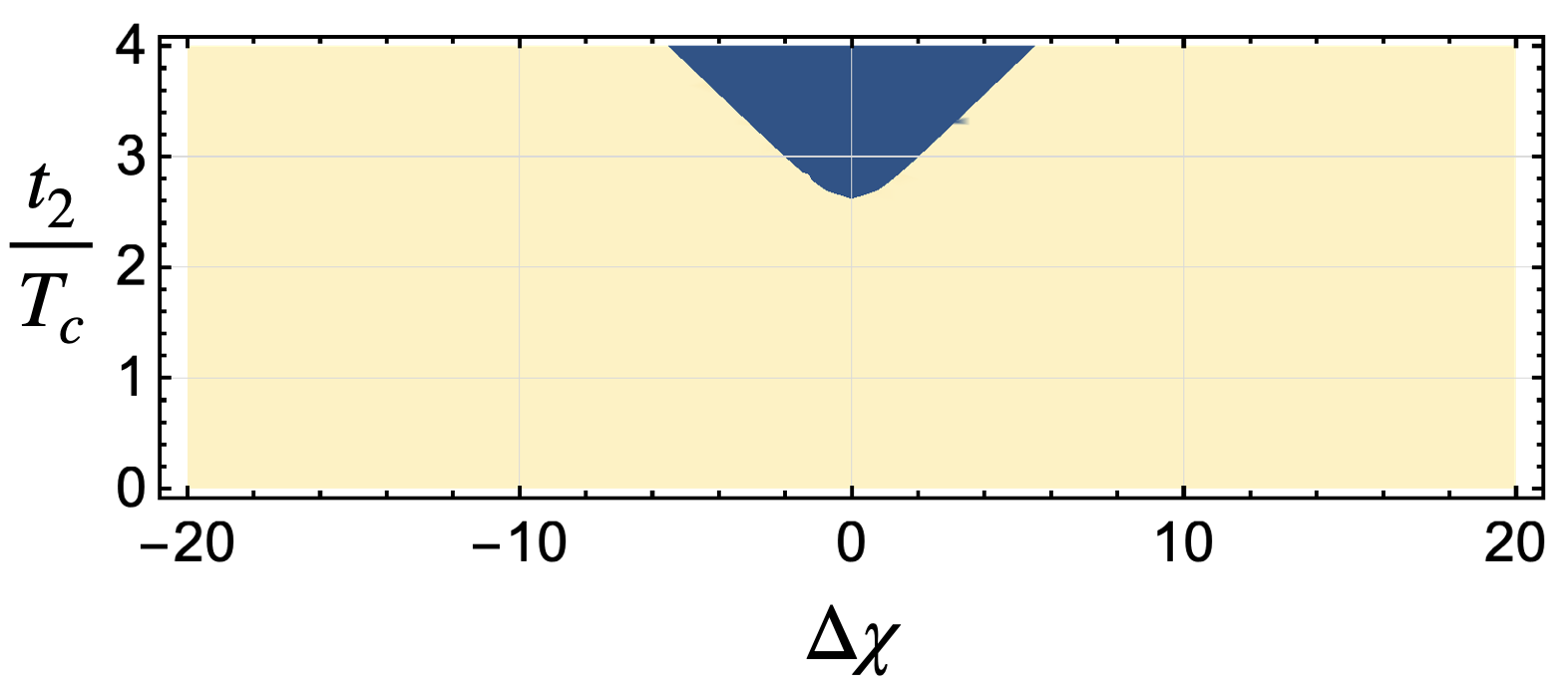}}
    \quad \quad
    \subfigure[$t_1=1.05T_c $]
    {\includegraphics[width=7.1cm]{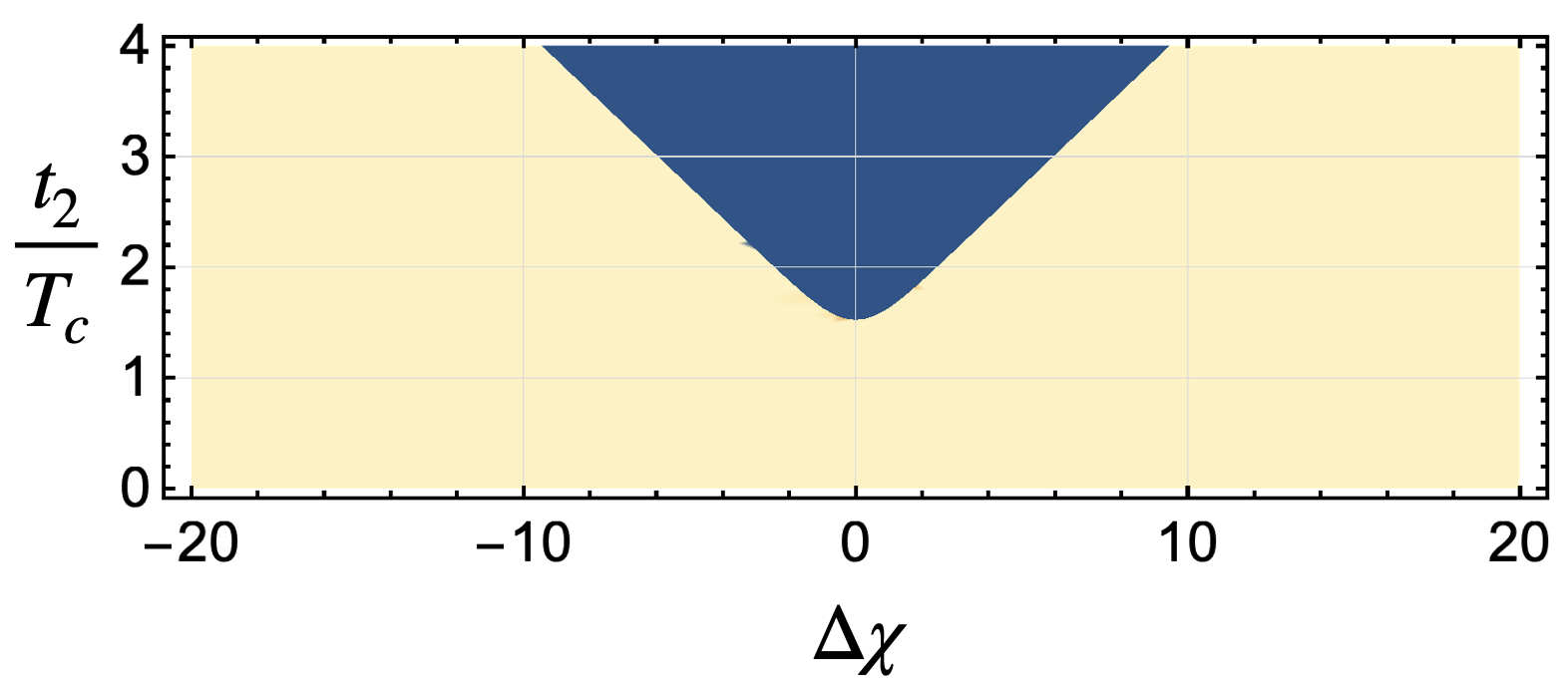}}\\
    \subfigure[$t_1=2T_c $] 
    {\includegraphics[width=7.1cm]{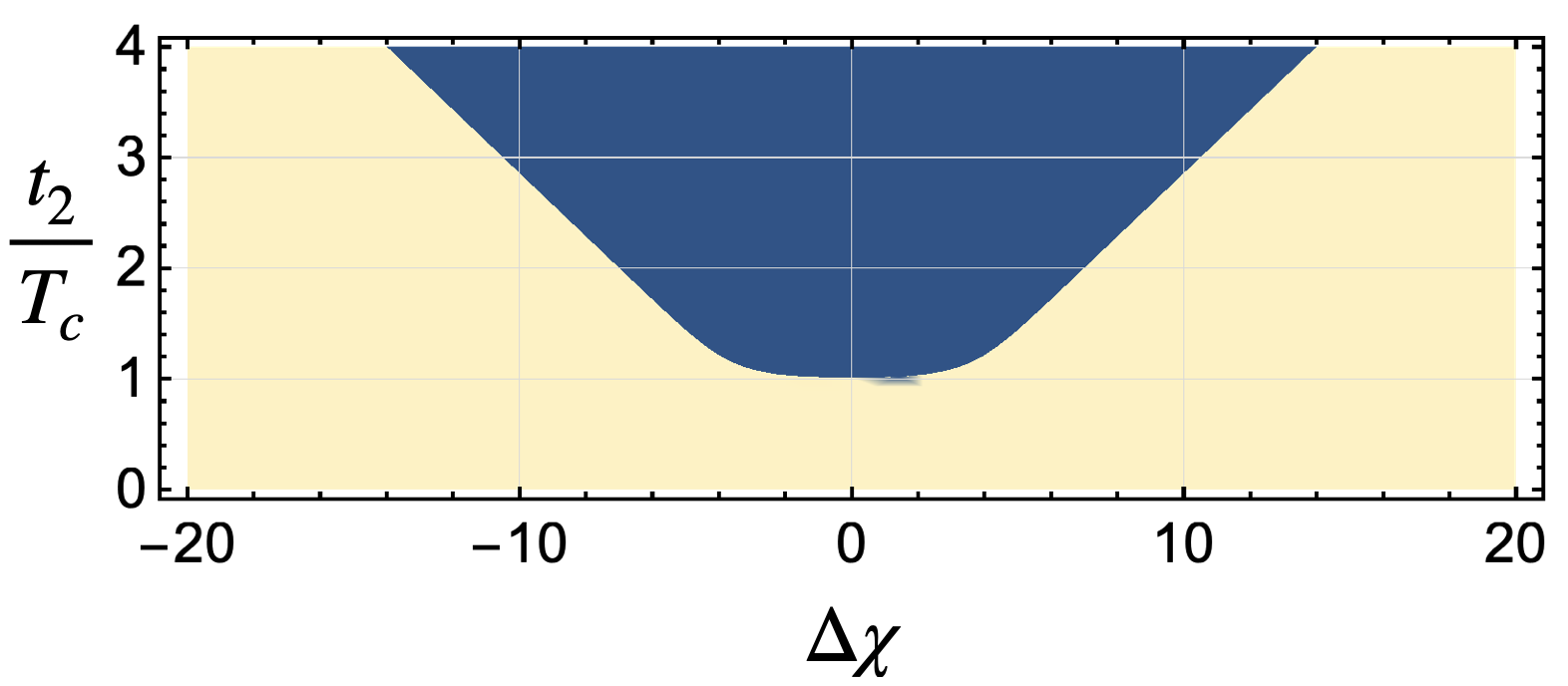}}
    \quad \quad
    \subfigure[$t_1=3T_c $]
    {\includegraphics[width=7.1cm]{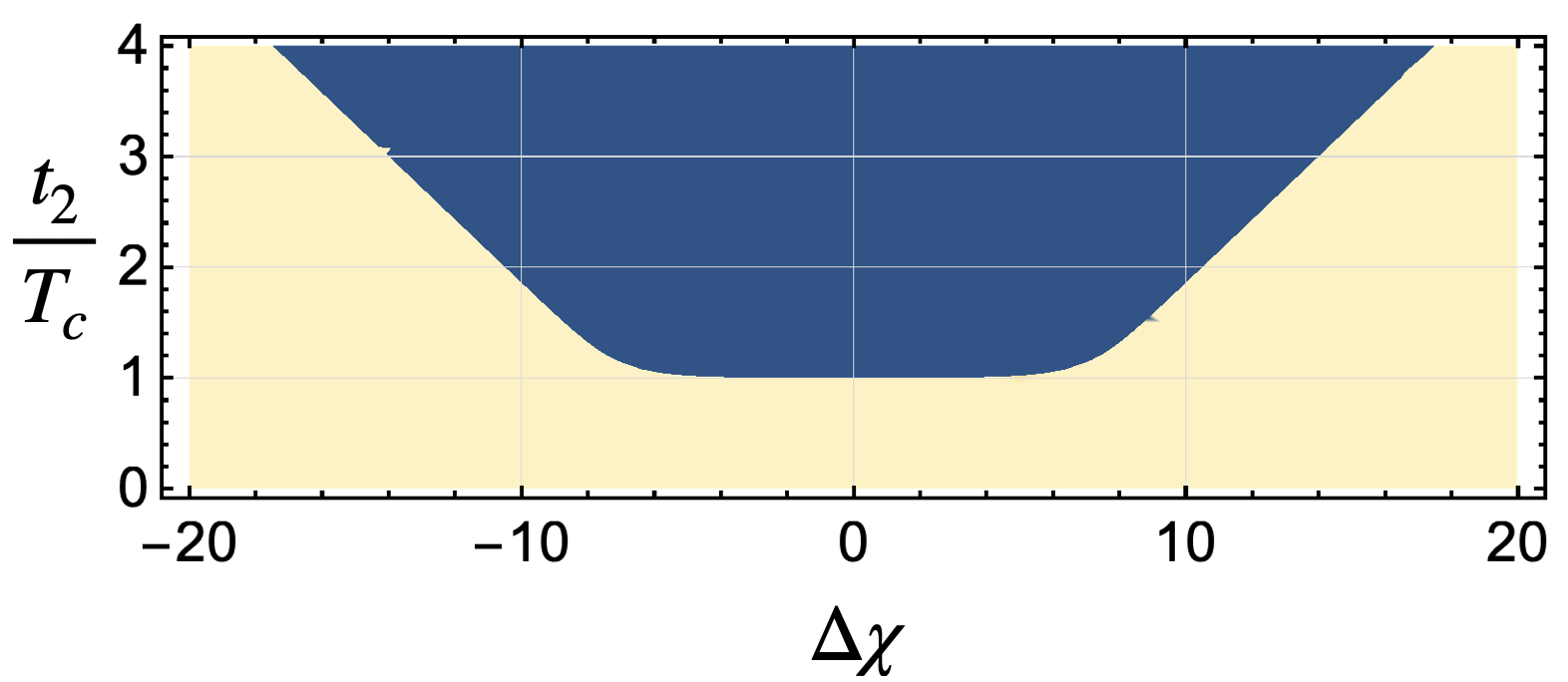}}
    \caption{Blue region (the light-cone like region) in the space of $(t_2/T_c,\Delta \chi)$ where the imaginary part of $C_\text{probe}$ takes non-zero values. The light-cone like region only appears for $t_2 \geq |T_c|$ and $|t_1| > |T_c|$. The commutator diverges when $\psi_L$ and $\psi_R$ are light-like separated (the boundary of the blue region), and its value decreases very quickly to zero as move to the region inside the blue region, taking non-zero values in the blue regions shown above. Here we set $t_0=0, \Delta_{\mathcal{O}}=1.4,$ and $G_N=1.$}
\label{TWC-hom}
\end{center}
\end{figure}

By writing $D=D_0+D_1 q$, and integrating \eqref{eq-Ctilde} with respect to $q$, we obtain
\be \label{Cprobe-hom}
 C_\text{probe}=-\alpha 2^{4 \Delta_{\psi}} b_{\psi}^{2}  \int d {\bf \tilde{x}_1} \frac{\Gamma(2\Delta_{\psi}) e^{(t_{1}-t_{2})\Delta_{\psi}}}{\left[ 2\left( e^{-t_{2}}\cosh{d({\bf \tilde{x}_{1},x_{2}})}+e^{t_{1}}\cosh{d({\bf \tilde{x}_{1},x_{1}})}\right)-D_{1} \right]^{2\Delta_{\psi}}}
 \ee
where we used the relation $C= e^{-i  \langle \mathcal{V} \rangle} \tilde{C}$, with $ \langle \mathcal{V}\rangle = D_0$. To study the behavior of commutator (\ref{eq-V}), we will consider the behavior of the correlator $C_{\text{probe}}$ whose imaginary part gives the original commutator $\langle \left[ \psi_{R}, e^{-i \mathcal{V}} \psi_{L} e^{i  \mathcal{V}} \right] \rangle = -2 i\, \text{Im} (C)$. We set $\chi_1=-\chi_2= \frac{\Delta \chi}{2}$ and write the geodesic distances as $d({\bf \tilde{x}_1,x_1})=\tilde{\chi_1}-\frac{\Delta \chi}{2}$ and $d({\bf \tilde{x}_1,x_2})=\tilde{\chi_1}+\frac{\Delta \chi}{2}$. In this case, the correlator (\ref{Cprobe-hom}) depends on the boundary parameters $(t_1,t_2,\Delta \chi)$ and on the scaling dimension $\Delta_{\psi}$ of the signal, as well as the information from the double trace deformation, which is encoded in $D_1$.
To investigate the traversability of the wormhole, we follow the method of \cite{almheiri2018escaping}. As the time interval between $t_1$ and $t_2$ increases, the denominator decreases and becomes negative. Thus, we can find the line where the denominator of the integrand in \eqref{Cprobe-hom} vanishes and the region where the commutator \eqref{eq-V} takes non-zero values. Inside this region, the wormhole becomes traversable. This region forms a sort of light-cone interior in the sense that the slope of its boundary is the speed of light.
In order to find this region,
 we fix the value of $t_1$ and study the behavior of the commutator as a function of $t_2$ and $\Delta \chi$. The result is shown in Fig.~\ref{TWC-hom}, in which the interior of the light-cone like region is shown in blue. In this region the commutator is non-zero for several values of $t_1$.


Consistently with \cite{almheiri2018escaping}, we observe that the commutator is zero if $t_2$ is smaller than the absolute value of a critical time scale which is given by $T_c = \log D_1 < 0$. Besides, the region of non-zero commutator only appears in a finite time if we choose $|t_1| >|T_c|$. The interpretation is that the signal should be early enough to be able to escape from the black hole and reach the right boundary. Interestingly, the critical time $T_c$ plays a role similar to the role played by the scrambling time $t_* \sim \log \frac{1}{G_N}$ in the behavior of OTOCs. In fact, for $t_0=0$, we can write the critical time as $T_c = -t_*+\log(h d_1)$, where we wrote the phase shift as $D_1 = h G_N d_1$, where $d_1$ is a function of $\Delta_{\mathcal{O}}$, and $d$. This shows that are calculation is barely consistent with the probe approximation, being valid in a time window of size $\delta t \sim \log(h d_1)$. We refer to \cite{almheiri2018escaping} for a more detailed discussion about this point.

The commutator diverges \footnote{This happens in the probe approximation. The result is not divergent when one considers the backreaction of the probe. See for instance \cite{Maldacena_2017,Couch_2020}.} at the boundary of the blue region in Fig.~\ref{TWC-hom} and decays very quickly to zero  as move to the region inside the light-cone, taking non-zero values in the blue region. As $t_1$ approaches $T_c$, with $|t_1|>|T_c|$, the blue region where the commutator is non-zero moves more and more to the future. The light-cone like structure appears because we are considering homogeneous shocks for the backreaction of the field excitations produced by operators $\mathcal{O}_R$ and $\mathcal{O}_L$. These light-cones will be replaced by butterfly cones once we consider localized shocks. This will be discussed in the next subsection.

\subsubsection{Localized perturbations} \label{sec-localized}
 In this section, we consider the case in which the double trace deformation is produced by local operators. For simplicity, we do not consider any dependence on the coordinates on the sphere $S^{d-2}$, but our operators depend on the hyperbolic coordinate $\chi$. 
 
 To simplify our calculations, let us consider the double trace deformation of the form (\ref{eq-doubletrace}) in which
\be
h(t',{\bf x'})= h \,\delta(t'-t_0) \delta({\bf x'},0)\,,
\ee
in such a way that (\ref{eq-D1}) becomes 
\be \label{eq-D1local}
D_{1}(\chi_1) = -\alpha\,h \, b_{\mathcal{O}}^{2} \frac{\Delta_{\mathcal{O}} \Gamma(2\Delta_{\mathcal{O}})}{2} \text{vol}(S_{d-2}) \int_{0}^{\infty} d  \chi_4  \frac{16 \pi G_N\sinh^{d-2}\chi_4  }{ (\cosh \chi_4 \cosh{t_0} )^{2 \Delta_{\mathcal{O}} + 1}}\frac{\,e^{-\mu |\chi_4-\chi_1|} }{d}\,,
\ee 
 where we use $f({\bf \tilde{x}_1,\tilde{x}_4 })=e^{-\mu d({\bf \tilde{x}_1,\tilde{x}_4 })}/d$ as the shock wave transverse profile for local perturbations\footnote{The transverse profile satisfies \eqref{eq-transverse profile}. In hyperbolic space, it is shown in \cite{Ahn:2019rnq}  that $\mu$ is related to the butterfly velocity as $v_B = 1/\mu$. We will see that besides characterizing OTOCs, $v_B$ also plays an important role in GJW's traversable wormhole setup.} and use $d({\bf \tilde{x}_1,\tilde{x}_4 })=|\chi_1-\chi_4|$.

 
 The factor of $e^{-\mu |\chi_4-\chi_1|}$ makes it hard to compute the integral analytically for general $d$, especially because of the $\chi_1$ dependence. However, for a given $d$, we can compute the integral analytically, and use it to numerically evaluate the correlator $\tilde{C}$ in (\ref{eq-Ctilde}). We can use the relation $C= e^{-i  \langle \mathcal{V} \rangle} \tilde{C}$ to obtain the correlator $C$ defined in (\ref{eq-C}). By performing the integral in $q$ in (\ref{eq-Ctilde}), and using that $ \langle \mathcal{V}\rangle = D_0$, we can write
 \be
 C_\text{probe}=-\alpha 2^{4 \Delta_{\psi}} b_{\psi}^{2}  \int d {\bf \tilde{x}_1} \frac{\Gamma(2\Delta_{\psi}) e^{(t_{1}-t_{2})\Delta_{\psi}}}{\left[ 2\left( e^{-t_{2}}\cosh{d({\bf \tilde{x}_{1},x_{2}})}+e^{t_{1}}\cosh{d({\bf \tilde{x}_{1},x_{1}})}\right)-D_{1}({\bf \tilde{x}_1})\right]^{2\Delta_{\psi}}},
 \ee
where $b_{\psi}=\frac{\pi c_{\psi}}{2^{\Delta_{\psi}} \Gamma(\Delta_{\psi})}$. 

When we set $d=2$ and evaluate $D_1$ analytically, the above formula is consistent with the result for a BTZ black hole obtained in \cite{almheiri2018escaping, Cornalba_2007}. To compare with lower dimensional cases ($d \leq 2$), we take ${\bf x_{2}=x_{1}}$ and $t_{0}=0$, and obtain
 \be \label{eq-Cprobenumerical}
 C_\text{probe}=-\alpha\, 2^{4\Delta_{\psi}} b_{\psi}^{2} \int d {\bf \tilde{x}_{1}} \frac{\Gamma(2\Delta_{\psi})}{ \bigg[4\cosh{\left( \frac{t_{1}+t_{2}}{2}\right)}\cosh{d({\bf \tilde{x}_{1},x_1})}+D_{1}({\bf \tilde{x}_1})\, e^{\frac{t_{2}-t_{1}}{2}}\bigg]^{2 \Delta_{\psi}}}\,.
 \ee
 The above result is exactly the same as the results for $d=2$ obtained in \cite{Maldacena_2017,almheiri2018escaping} once we implement the replacement $D_1(\tilde{\bf x}_1) \rightarrow D_1$ which is necessary because we are considering a local perturbation, i.e., $h(t',{\bf x'}) \propto \delta({\bf x'},0)$.
 
 We now discuss the conditions under which traversability is optimal. In order to do that, one can define a `sweet spot' where traversability is optimal \cite{Maldacena_2017,Couch_2020}, which can be determined by the points where the commutator is maximal. To find the behavior of sweet spot, we will consider the maximal value of $C_{\text{probe}}$ and use the relation $\langle \left[ \psi_{R}, e^{-i \mathcal{V}} \psi_{L} e^{i  \mathcal{V}} \right] \rangle = -2 i\, \text{Im} (C)$. We put the double trace deformation at the origin of the coordinate system and at $t_0=0$, i.e., $h(t,\chi)=h\, \delta(t)\, \delta(\chi)$, and we choose the signal coordinates as $\chi_1=\chi_2=X$ and $t_2=-t_1=T$. For simplicity, we do not consider any dependence on the coordinates ${\bf x} \in S_{d-2}$. With the above definitions, the correlator (\ref{eq-Cprobenumerical}) becomes
\be 
C_\text{probe}(T,X)=-\alpha 2^{4 \Delta_{\psi}} b_{\psi}^{2} \text{vol}(S_{d-2}) \int d {\tilde{\chi}_1} \frac{\Gamma(2\Delta_{\psi}) }{\left[ 4\cosh{( \tilde{\chi}_{1}-X)}+e^{T} D_{1}({\tilde{\chi}_1})\right]^{2\Delta_{\psi}}}\,,
\label{CprobeInt}
\ee
where $D_1(\tilde{\chi}_1)$ is given by (\ref{eq-D1local}). 

\begin{figure}
\begin{center}
    \subfigure[Zero lines of the denominator of (\ref{CprobeInt}) for fixed values of $X$. The red dots indicate the points that have a dominant contribution to the integration.]
    {\includegraphics[width=7cm]{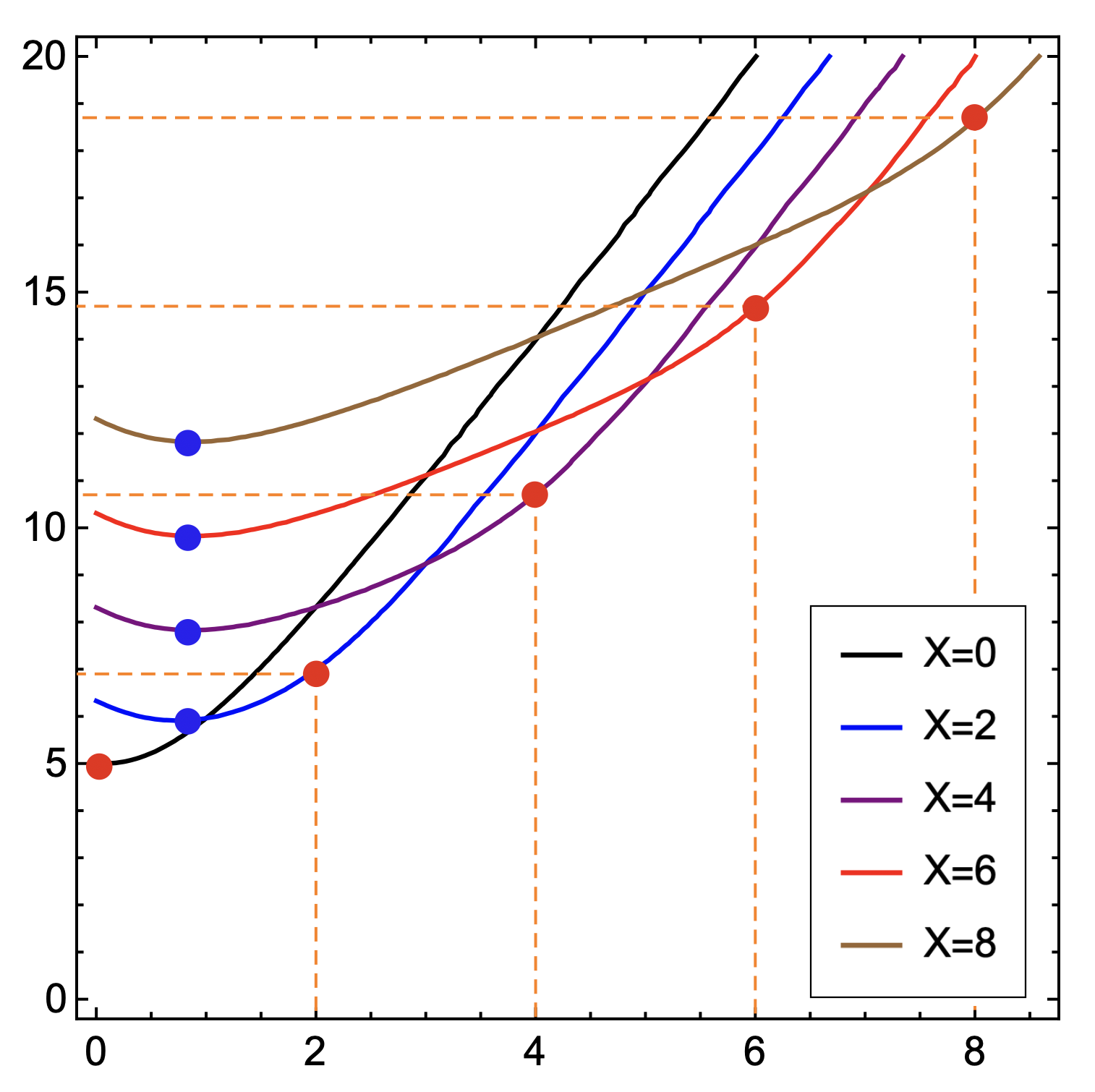}}
    \quad \quad
    \subfigure[The blue lines denote curves of zeros. The orange lines have slop of $2=1/v_{B}$ (solid) and $1=1/c$ (dashed).] 
    {\includegraphics[width=7cm]{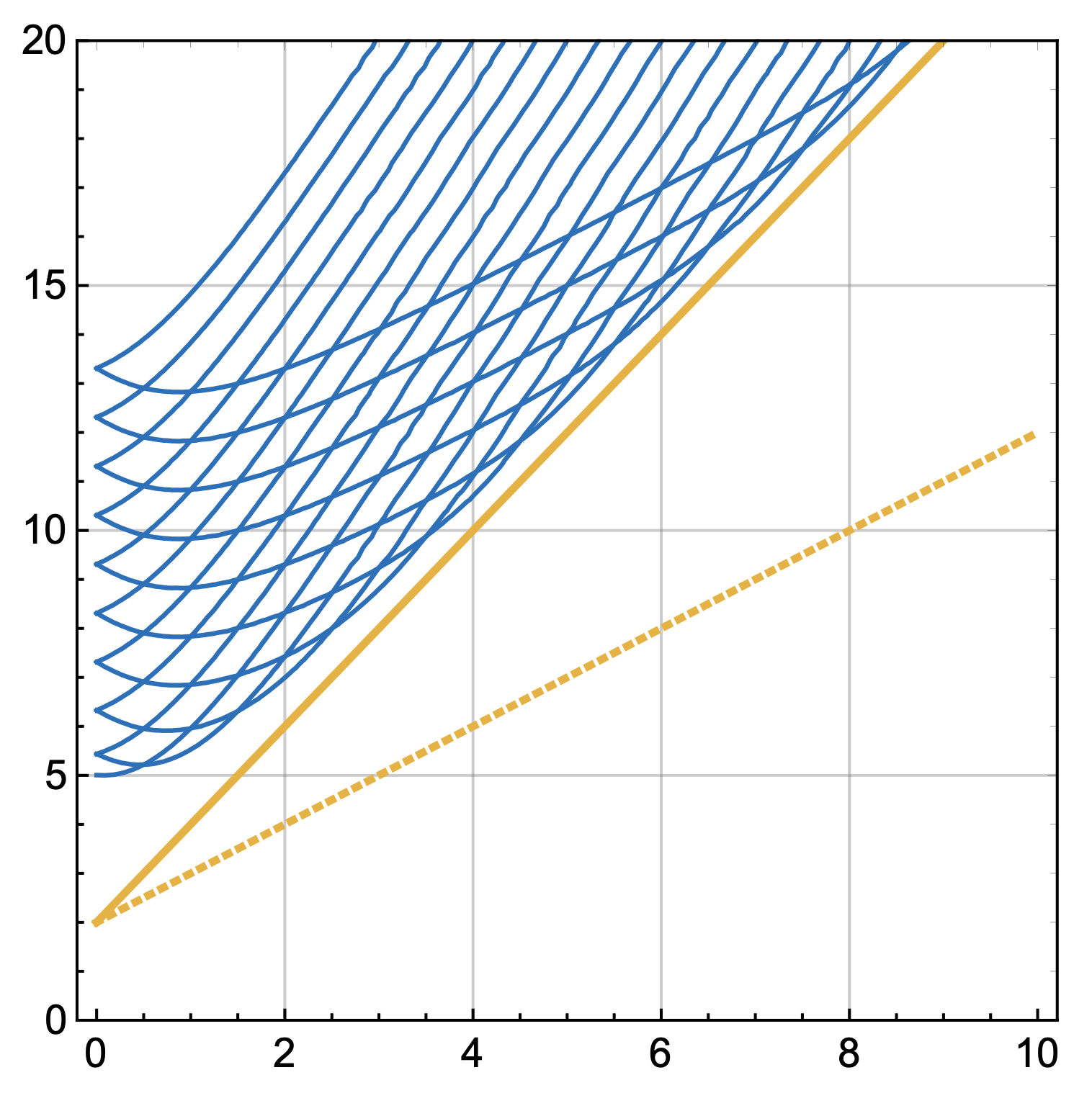}}
\put(-102,-4){$\tilde{\chi}_{1}$}
\put(-332,-4){$\tilde{\chi}_{1}$}
\put(-432,100){$T$}
\put(-207,100){$T$}
       
    \caption{Curves along which the denominator of (\ref{CprobeInt}) vanishes. Here, we consider a perturbation localized at $\chi_0 = 0$ and $t_0=0$, and we set $d=3$, $\Delta_\mathcal{O} = 1.4$ and $G_N=1$.}
\label{TWC-loc}
\end{center}
\end{figure}

Since it is hard to obtain precise results by numerically integrating (\ref{CprobeInt}),
we study the region inside which the commutator is non-zero by finding the zeros of the denominator in (\ref{CprobeInt}). More specifically, for a given $X$ we find the curves $T = F(\tilde{\chi}_1)$  such that
\be \label{eq-zeroline}
4\cosh{( \tilde{\chi}_{1}-X)}+e^{T} D_{1}({\tilde{\chi}_1})=0\,.
\ee
We plot the curves $T = F(\tilde{\chi}_1)$ for different values of $X$ in Fig.~\ref{TWC-loc}.

For a given fixed value of $T$ and $X$, the correlator $C_\text{probe}$ is obtained by performing an integral in $\tilde{\chi}_1$. We numerically observe that, for each $X$, the integral \eqref{CprobeInt} takes complex values in the region $T \geq F(\tilde{\chi}_1)$, and it is real for $T < F(\tilde{\chi}_1)$. That implies that the commutator $\langle \left[ \psi_{R}, e^{-i \mathcal{V}} \psi_{L} e^{i  \mathcal{V}} \right] \rangle = -2 i\, \text{Im} (C_\text{probe})$ takes non-zero values in the region $T \geq F(\tilde{\chi}_1)$. Inside this region, traversability is possible when the commutator takes order one values.

We now would like to find the region in the space of $(T,X)$ inside which traversability is optimal.
For a given $X$, the minimum value of $T$ at which the commutator is non-zero is indicated by the blue dots in Fig.~\ref{TWC-loc}. 
The optimal condition for traversability, however, happens at a slightly later time. In fact, we numerically observe that the dominant contribution to the integral (\ref{CprobeInt}) comes from the region at which
 $\tilde{\chi}_1 \approx X$, and the maximal value of commutator is obtained around the point $(\tilde{\chi}_1,T) \approx (X,F(X))$, indicated by the red dots in Fig.~\ref{TWC-loc}. Note that the red dots roughly indicate the point after which the curve $T = F(\tilde{\chi}_1)$ becomes a straight line for each $X$. By considering the curves $T = F(\tilde{\chi}_1)$ for increasing values of $X$, we observe that the collection of red dots forms a curve that approaches the orange line in Fig.~\ref{TWC-loc}. Considering $\Delta_{\mathcal{O}} \gg 1$, we numerically check that the slope of the orange line approaches $1/v_B$. We also observe that the gap near $\tilde{\chi}_1=0$ between the orange line and the curves $T = F(\tilde{\chi}_1)$ shrinks for $\Delta_{\mathcal{O}} \gg 1$. That basically implies that along the orange line we have $T \approx \frac{1}{v_B} X + T^*$, where $T^*$ is a constant. This shows that the ``sweet spot'' for traversabilty is determined by the butterfly speed, $v_B$, and by the time scale $T^*$. We will see in the following that $T^*$ is closely related to the scrambling time.

To better understand the above numerical observations, we also evaluate $C_\text{probe}(T,X)$ using a saddle point approximation. In the limit, $1 \ll \Delta_{\psi} \ll \Delta_{\mathcal{O}}$, one can find that the integral (\ref{CprobeInt}) is dominated by the region in which $\tilde{\chi}_1 \approx X$. As a result, the correlator is well approximated by the following formula
\be \label{eq-SaddleC}
C_\text{probe} \sim  \frac{1}{\left[ 4 + e^{T} D_{1}(X)\right]^{2\Delta_{\psi}}}\,.
\ee
In the limit $1 \ll \Delta_{\mathcal{O}} $ taken above, the integral (\ref{eq-D1local}) giving $D_1$ is dominated by the region in which $\tilde{\chi}_4 \approx 0$, which leads to the simple result
 \be \label{eq-D1simple}
D_{1}(X) \approx - \gamma_{\mathcal{O}} \,h\,G_{N} e^{-(d-1)X}\,.
\ee
where $\gamma_{\mathcal{O}}=8 \pi  \alpha b_{\mathcal{O}}^2 \Delta_{\mathcal{O}} \Gamma\left(2\Delta_{\mathcal{O}}\right)$. The correlator can then be written as
\be \label{eq-SaddleC2}
C_\text{probe} \sim  \frac{1}{\left[ 1 - e^{T-T^*-(d-1)X}\right]^{2\Delta_{\psi}}}\,.
\ee
where $T^*=\log \left( {\frac{4}{\gamma_{\mathcal{O}}\,h\, G_N}} \right)$. From (\ref{eq-SaddleC2}) , one can see that $C_\text{probe}$ behaves like an OTOC, being characterized by
a unity Lyapunov exponent and butterfly speed given by $v_B =\frac{1}{d-1}$. The only difference is that now the critical time $T^*$ is not precisely equal to the scrambling time, because it also involves the coupling $h$. Note that $C_\text{probe}$ diverges along the line $T=T^*+ X/v_B$. This line reproduces the orange line in Fig.~\ref{TWC-loc} in the limit $\Delta_{\mathcal{O}} \gg 1$.

Therefore, from (\ref{eq-SaddleC2}) it is clear that the butterfly speed $v_B =\frac{1}{d-1}$ plays an important role in the behavior of the commutator, and the `sweet spot' for traversability is indeed controlled by $v_B$ when both the signal and the deformation are produced by local operators.
The butterfly speed of the sweet spot defines a butterfly cone inside which the commutator is non-zero.

In the case of a BTZ black hole, the butterfly cone is indistinguishable from a light-cone \cite{almheiri2018escaping, Couch_2020}. However, for $d > 2$ the butterfly speed is smaller than the speed of light (for Rindler-AdS $v_B=\frac{1}{d-1}$), and the cones are clearly distinguishable. Here, our result shows that $v_B$ plays a very important role in holographic teleportation protocols, having the same relevance that it has in controlling the spatial behavior of OTOCs. To our knowledge, this provides the first example in which the optimal condition for traversability is controlled by the butterfly speed, with $v_B < 1$.\footnote{ In principle, one can also use the point splitting method to study traversability considering localized perturbations. However, in this case we cannot set $\mathbf{x}=\mathbf{x'}=0$ in \eqref{G1UU}, and this makes calculation more complicated.}

\subsection{Beyond the probe approximation} \label{sec-backreaction}

So far, we have assumed the probe approximation. In this section, we consider the effect of the backreaction of the signal produced by the boundary operator $\psi$. Contrary to the double trace deformation, which creates a negative-energy shock wave that opens the wormhole, the signal creates a positive-energy shock wave that closes the wormhole. The opening of the wormhole is diagnosed by  a violation of the ANEC, which makes $\Delta V <0$. The backreaction of the signal introduces a positive contribution that makes $\Delta V$ less negative, i.e., $\Delta V \rightarrow \Delta V_\text{back}$, with $|\Delta V_\text{back}| \leq |\Delta V|$. 

Studying the backreaction effect in the proper way requires us to consider coupled quantum fields and re-compute the stress tensor \cite{Freivogel_2020}. Instead of the full consideration of this effect, we will consider an heuristic method and follow the approach adopted in \cite{Caceres_2018}, which consists in setting the parametric behavior that corresponds to the backreaction effect, then introducing an extra probe particle that experiences the shift $\Delta V_\text{back}$.

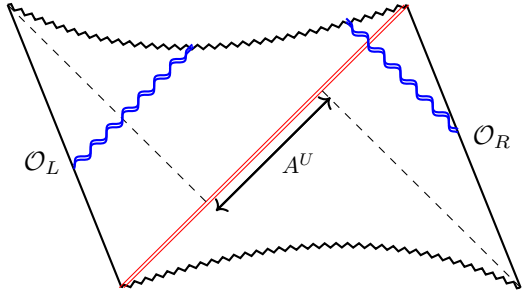
\begin{figure}
\centering

\begin{tikzpicture}[scale=1.5]
\draw [thick]  (0,0) -- (-1,2.5);
\draw [thick]  (3.5,0) -- (2.5,2.5);

\draw [thick,decorate,decoration={zigzag,segment length=1.5mm, amplitude=0.3mm}] (-1,2.5) .. controls (0,2) 
and (1,2) .. (2.5,2.5);
\draw [thick,decorate,decoration={zigzag,segment length=1.5mm,amplitude=.3mm}]  (0,0) .. controls (1.5,.5) and (2.5,.5) .. (3.5,0);

\draw [red]  (0.02,0) -- (2.52,2.5);
\draw [red]  (-0.02,0) -- (2.48,2.5);
\draw [dashed]  (-1,2.5) -- (0.75,0.75);
\draw [dashed]  (3.5,0) -- (1.75,1.75);

\draw [thick,->]  (0.83,0.68) -- (1.83,1.68);
\draw [thick,->]  (1.83,1.68) -- (0.83,0.68);

\draw [thick,blue,decorate,decoration={snake,segment length=3mm,amplitude=0.5mm}]  (2.95,1.4) -- (1.97,2.38);
\draw [thick,blue,decorate,decoration={snake,segment length=3mm,amplitude=0.5mm}]  (2.95,1.37) -- (1.97,2.35);

\draw [thick,blue,decorate,decoration={snake,segment length=3mm,amplitude=0.5mm}]  (-0.42,1.08) -- (0.62,2.15);
\draw [thick,blue,decorate,decoration={snake,segment length=3mm,amplitude=0.5mm}]  (-0.42,1.05) -- (0.63,2.13);

\node[scale=0.8,align=center] at (1.55,1.1) {$A^U$};

\end{tikzpicture}
\vspace{0.1cm}
\put(-190,45){\rotatebox{0}{\small $\mathcal{O}_L$}}
\put(-22,55){\rotatebox{0}{\small $\mathcal{O}_R$}}

\caption{\small Positive-energy shock wave geometry produced by the backreaction of the signal, which is introduced by the operator $\psi_L$. The negative-energy shock wave produced by the operator $\mathcal{O}_R$ suffers a time delay as it crosses the shock.}
\label{fig-backreaction}
\end{figure}

For simplicity, we consider a signal traveling along the $V=0$ horizon. The signal produces a shock wave geometry that affects the negative-energy shock produced by the operator $\mathcal{O}_R$, as shown in Fig.~\ref{fig-backreaction}.  
The opening of the wormhole is measured by the parameter $A^{U}$, which is related to the total momentum of the signal as
\be \label{eq-AU}
A^{U}= \frac{16 \pi G_N }{d-1}\, q^{\text{tot}},
\ee
where we consider a signal that is homogeneous along the transverse coordinates. The effect of this shock wave is to produce a time-delay in the negative-energy shock produced by $\mathcal{O}_R$, which can be incorporated by changing its wave function as
\be 
\Psi_{\mathcal{O}_R} \rightarrow e^{i A^{U} p}\Psi_{\mathcal{O}_R}\,,
\ee
in (\ref{eq-expD}). Note that this change can be incorporated by replacing the phase shift as
\be
\delta \rightarrow \delta_\text{back}=\delta +A^{U} p\,,
\ee
with $\delta$ given by
\be
\delta = \frac{16 \pi G_N}{d-1} \,q \,p,
\ee
where $q$ is the total momentum of the probe particle in the $U$ direction. Using (\ref{eq-AU}), we can write
\be
\delta_\text{back} = \frac{16 \pi G_N}{d-1} (q+q^\text{tot})\, p\,.
\ee
With the above formulas, we can then rewrite $D$ in (\ref{eq-expD2}) as
\be \label{eq-expD3}
 D= \alpha  \frac{\pi^{2} c_{\mathcal{O}}^{2} 2^{2\Delta_{\mathcal{O}}}}{\Gamma(\Delta_{\mathcal{O}})^{2}} \int d p  d t'\, d {\bf x'} p^{2\Delta_{\mathcal{O}}-1} e^{i 4 p \cosh{d({\bf x',0})} \cosh{t'}} e^{-i \pi \Delta_{\mathcal{O}}} e^{i \frac{16 \pi G_N}{d-1} (q+q^\text{tot})\, p},
 \ee
where we set ${\bf \tilde{x}_4}=0$ and do not integrate over ${\bf \tilde{x}_4}$ because we are considering a homogeneous perturbation. Integrating (\ref{eq-expD3}) in $p$ we obtain
\be \label{eq-Dback1}
D = \alpha \,2^{4 \Delta_{\mathcal{O}}} b_{\mathcal{O}}^{2} \int d t' d {\bf x'} \frac{\Gamma(2 \Delta_{\mathcal{O}}) h(t',{\bf x'})}{\big[ 4 \cosh{d({\bf x',0}) \cosh{t'} + \frac{16 \pi G_{N}}{d-1} (q+q^\text{tot}) } \big]^{2 \Delta_{\mathcal{O}}}}\,.
\ee
Now, to obtain the shift $\Delta V_\text{back}$ suffered by the probe particle, we expand (\ref{eq-Dback1}) for small values of $q$ and extract the coefficient of the linear order term in $D=D_{0}+D_{1} q$. By using (\ref{eq-ANED1}), we find
\be \label{eq-deltavback1}
\Delta V_\text{back} = - \Delta_\mathcal{O} \frac{16 \pi G_N}{d-1} \alpha \,2^{4 \Delta_{\mathcal{O}}} b_{\mathcal{O}}^{2} \int d t' d {\bf x'} \frac{\Gamma(2 \Delta_{\mathcal{O}}) h(t',{\bf x'})}{\big[ 4 \cosh{d({\bf x',0}) \cosh{t'} + \frac{16 \pi G_{N}}{d-1} q^\text{tot} } \big]^{2 \Delta_{\mathcal{O}}+1}}\,.
\ee
To evaluate the above expression, we consider an instantaneous perturbation $h(t',{\bf x'})= \delta(t'-t_0)$ with operators that are homogeneous on $S_{d-2}$. We obtain
\be \label{eq-Dback2}
\Delta V_\text{back} = -\frac{4 \pi G_N}{d-1} \Delta_\mathcal{O} \alpha \, b_{\mathcal{O}}^{2} \frac{\text{vol}(S_{d-2})}{(\cosh t_0)^{2 \Delta_\mathcal{O}+1}}\int_{0}^{\infty} d \chi' \frac{ \sinh^{d-2} \chi'\, \Gamma(2 \Delta_{\mathcal{O}})}{\big[ \cosh{\chi'}  + \frac{4 \pi G_{N} q^\text{tot}}{(d-1)\cosh t_0}  \big]^{2 \Delta_{\mathcal{O}}+1}}\,,
\ee
which can be integrated to give
\be \label{eq-deltavback2}
\Delta V_\text{back} = d_\mathcal{O} \frac{\Gamma(2\Delta_\mathcal{O}-d+3)\Gamma(\frac{d-1}{2})}{2 \Gamma(2\Delta_\mathcal{O}-\frac{d-5}{2})} \frac{F_{1}\bigg(2\Delta_\mathcal{O}-d+3,2\Delta_\mathcal{O}+1,\frac{3-d}{2},2\Delta_\mathcal{O}-\frac{d-5}{2},-\frac{4 \pi G_N q^\text{tot}}{(d-1)\cosh{t_{0}}},-1\bigg)}{ (\cosh{t_0})^{2\Delta_\mathcal{O}+1}}\,,
\ee
where $F_1$ is the Appell hypergeometric function and $d_\mathcal{O}=-\frac{8 \pi G_N}{d-1} \Delta_\mathcal{O} \alpha \, b_{\mathcal{O}}^{2} \text{vol}(S_{d-2})$. The above formula is valid for $2 \Delta_\mathcal{O}>d-3$, and $d>1$, which is always true for the cases we consider. Note that our result for $\Delta V_\text{back}$ corresponds to a correction to the wormhole opening $\Delta V $ obtained in Sec.~\ref{sec-homogeneous} in the probe limit and for homogeneous perturbations.

\begin{figure}
\begin{center}
\begin{tabular}{cc}
\setlength{\unitlength}{1cm}
\hspace{-0.9cm}
\includegraphics[width=7.5cm]{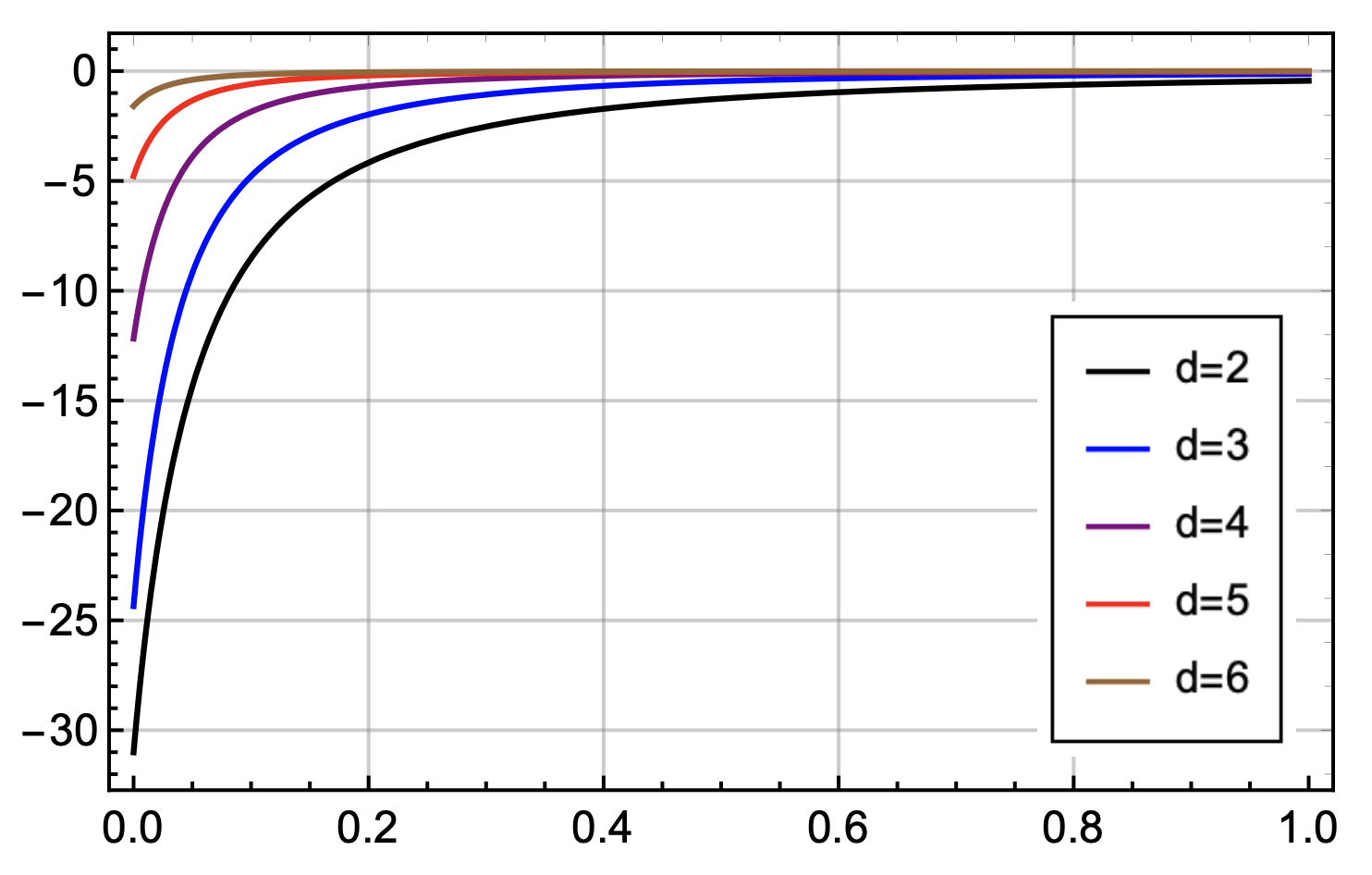}
\qquad \qquad &
\includegraphics[width=7.5cm]{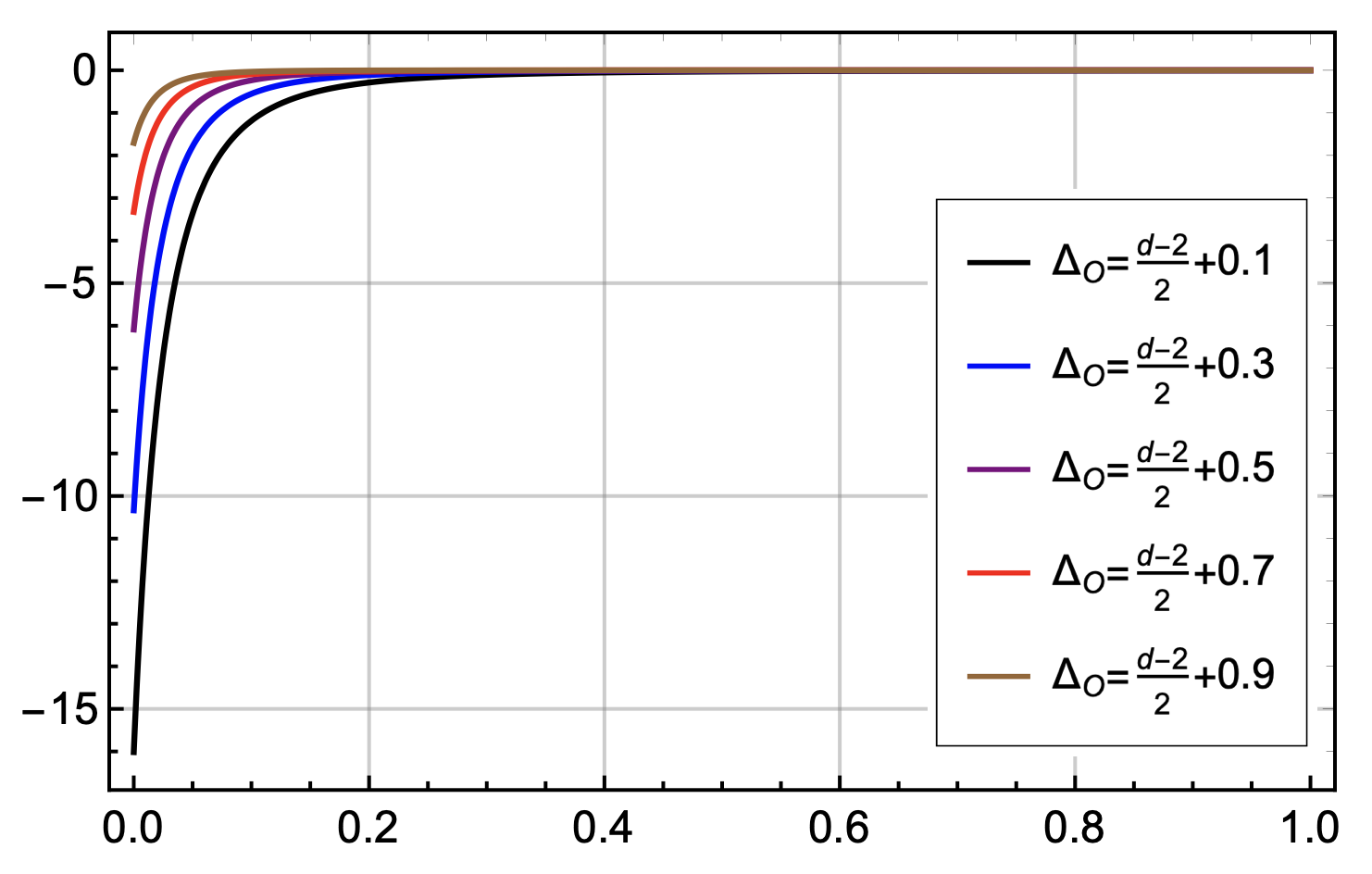}
\qquad
\put(-108,-10){$q^\text{tot}$}
\put(-358,-10){$q^\text{tot}$}
\put(-480,50){\rotatebox{90}{$\Delta V_\text{back}$}}
\put(-225,50){\rotatebox{90}{$\Delta V_\text{back}$}}
\end{tabular}
\end{center}
\caption{ $\Delta V_\text{back}$ versus the total momentum of the signal. As the total momentum of the signal increases, its backreaction becomes stronger and closes the wormhole. On the left panel, we show the plots for several values of $d$. Here, we set $\Delta_{\mathcal{O}} = \frac{d-2}{2}+0.5$ for each $d$. On the right panel, we show the dependency on the conformal dimension when the $d$ is fixed. As an example, we draw the $d=3$ case, but other values of $d$ lead to the same qualitative behavior. Here we use $t_{0}=0$.}
\label{fig-back}
\end{figure}

The opening of the wormhole as seen by the probe particle now depends on the total momentum of the signal $q^\text{tot}$, and in fact $\Delta V_\text{back}$ approaches zero as we increase $q^\text{tot}$. This is seen in Fig.~\ref{fig-back}. This backreaction effect has important consequences on the amount of information that can be sent through the wormhole. This will be discussed in the Sec.~\ref{sec-back-info}

The result of Fig.~\ref{fig-back} can also be obtained using the point splitting method, as explained in \cite{fallows2020making}. This paper considers an $AdS_{2}$ geometry which is closely related to the geometry considered in \cite{Maldacena_2017}. Indeed, by taking the correspondence between the formula in \cite{fallows2020making} and our formula (\ref{eq-stress}), one might check that the backreaction effect from the signal can be incorporated by introducing a shift $\delta U$ in the $U$ variable in (\ref{eq-A10}). We will leave a detailed investigation of this for future work. 

\section{Bound on information transfer in higher dimensions} \label{sec-bound}
 
 In the previous section, we show that the wormhole connecting the two Rindler wedges of AdS becomes traversable once we turn on a double trace deformation coupling the two asymptotic boundaries. In this configuration, information can be transferred through the wormhole, which can be diagnosed by a non-zero two-sided correlation function.
 In this section, we compute parametric bounds on the amount of information that can be transferred through the wormhole. The signal/message will be described by a positive-energy shock wave in the bulk that interacts with a negative-energy shock wave introduced by the double trace deformation.

 \subsection{Overview}
 
 First, we briefly review the derivation of information transfer bounds for lower dimensional ($d \leq 2$) black holes ~\cite{Freivogel_2020,Maldacena_2017,Caceres_2018}. We then extend these ideas to higher dimensional ($d \geq 2$) setups and compare them with the analysis presented in \cite{Freivogel_2020}.
 
 The bound on information transfer appears because the backreaction of the signal closes the wormhole. The idea is as follows. First, the double trace deformation introduces a negative-energy shock wave in the bulk that opens the wormhole, allowing probe particles to cross from one side of the geometry to the other side. Therefore, information can be transferred through the wormhole by sending several particles, each one corresponding to a bit of information.
 However, a signal containing too many particles might have a backreaction on the geometry, which is described by a positive-energy shock wave geometry that closes the wormhole. This effect limits the amount of information that can be transferred through the wormhole - if we send too many bits, the wormhole closes.

Although it is hard to precisely derive a bound on the information transfer, it is possible to derive parametric bounds, in which one cares about the parametric scaling of the bound and ignores constant factors. Let $p_V^\text{tot}$ be the total momentum of a signal containing $N_\text{bits}$ particles, each one with momentum $p_V^\text{each}$, in such a way that we can write:
\be \label{eq-Nbits}
N_\text{bits}=\frac{p_V^\text{tot}}{p_V^\text{each}}\,.
\ee
We can now derive a bound on $N_\text{bits}$ using the uncertainty principle and requiring that the signal `fits' in the opening of the wormhole. The uncertainty principle states that:
\be
p_V^\text{each} \Delta V_\text{each} \gtrsim 1\,.
\ee
Let us now assume that the double trace deformation opens the wormhole by an amount $\Delta V$. In order for the signal wave function to pass through the wormhole, we need that 
\be
\Delta V_\text{each} \leq | \Delta V |\,,
\ee
which implies that
\be \label{eq-uncertainty}
p_V^\text{each}  \gtrsim \frac{1}{\Delta V_\text{each}} \geq \frac{1}{| \Delta V |}\,.
\ee
Combining (\ref{eq-uncertainty}) with (\ref{eq-Nbits}) we find
\be \label{eq-Npvdeltav}
N_\text{bits}  \lesssim p_V^\text{tot} |\Delta V|\,.
\ee
Finally, we require that the backreaction of the signal is small, in such a way that it does not destroy the negative-energy shock wave geometry. The precise form of this probe approximation depends on some details of the system in consideration, and whether the signal is localized or not. For example, for a BTZ black hole, the probe approximation implies \cite{Caceres_2018, Freivogel_2020}
\be
\frac{G_N p_V^\text{tot} }{r_0} \ll 1\,,
\ee
where $r_0$ is the horizon radius of the black hole in Schwarzschild coordinates. Using the explicit form of the ANE for the BTZ black hole $\Delta V  \sim \frac{G_N h}{\ell} K$ \cite{Freivogel_2020}, where $\ell$ is the AdS radius, one finds
\be \label{eq-boundN}
N_\text{bits} \lesssim h \frac{r_0}{\ell} K,
\ee
where we consider the double trace deformation (\ref{eq-doubletrace}) with $K$ fields with $h(t',{\bf x'})= h\, \theta(t'-t_0)$ and the signal that is homogeneous in the transverse coordinates. The number of fields $K$ should also be constrained for this construction to be reliable. The authors of \cite{Freivogel_2020} argue that $K \lesssim \frac{\ell}{G_N}$, which combined with (\ref{eq-boundN}) gives
\be
N_\text{bits} \lesssim h\,\frac{r_0}{G_N} \approx h\, S_\text{BH}\,,
\ee
which implies that the information transfer is bounded by the black hole entropy $S_\text{BH}$.

The precise parametric form of this bound changes in higher dimensional setups, and it also depends on whether the signal is localized or not. We will dive into these details in the next subsections.


 \subsection{Homogeneous shocks}

To derive parametric bounds on information transfer, we will use (\ref{eq-Npvdeltav}) with the explicit form of $\Delta V$, which in general depends on $\Delta_\mathcal{O}$ and $d$, combined with the probe approximation for the signal. To impose the probe approximation, we will require that the classical action of the signal on the background of the negative-energy shock wave is small. That is equivalent to say that the phase shift $\delta$ that controls the interaction of the signal with the shock wave produced by the operator $\mathcal{O}$ is small.

For simplicity, in this section we only consider shock waves that are homogeneous in the transverse space, i.e., they do not depend on the transverse coordinates ${\bf x} \in \mathcal{H}_{d-1}$.
Let us start by describing the shock wave geometry produced by the negative energy. We take the stress energy tensor of the shock as (see, for instance \cite{Shenker:2014cwa,Ahn:2019rnq}):
\be
T_{UU} =\frac{q_{U}}{r_0^{d-1}}\delta(U)\,.
\ee
where $r_0$ is the horizon radius, which we reintroduce for later convenience, and $q_U$ is the total momentum associated with the shock. The corresponding backreaction on the geometry is simply obtained by the replacement
\be
ds^2 \rightarrow ds^2+ h_{UU}^{-} dU^2\,, \,\,\,h_{UU}^{-}= \frac{16 \pi G_N}{r_0^{d-3}} q_{U}\,\delta(U)\,f({\bf x-x'}),
\ee
where $ds^2$ is the unperturbed geometry (\ref{eq-metric-kruskal}), and $f({\bf x})$ is the shock wave transverse profile, which satisfies the following equation
\be\label{eq-transverse profile}
\left( \Box_{\mathbb{H}_{d-1}} - \frac{2 \pi}{\beta} r_{0} (d-1)\right) \, f({\bf x})= 1.
\ee
Since $T_{UU}$ has no ${\bf x}$ dependence, we look for a solution in which $f$ is constant. We obtain
\be
f= \frac{\beta }{2\pi (d-1)r_0} \,.
\ee
We can finally write:
\be \label{eq-hUU}
h_{UU}^{-}= \frac{16 \pi G_N}{r_0^{d-3}} q_{U}\,\delta(U)\, \frac{1}{\mu}\,,\,\,\,\, \mu \equiv (d-1)r_0^2 \,.
\ee
We now consider the stress energy tensor of the signal, which we also take as homogeneous:
\be
T_{VV}^{+}=\frac{p_{V}}{r_0^{d-1}} \delta(V)\,.
\ee
We can then compute the phase shift of the collision between the two shocks as~\cite{Shenker:2014cwa}
\be \label{eq-classical action}
\delta = S_\text{classical}=\frac{1}{2} \int d^{d+1}x\sqrt{-g}\,
h_{UU}^{-}\, T^{UU}_{+}= \text{vol}(\mathbb{H}_{d-1}) \frac{4 \pi G_N}{r_0^{d-3}} \frac{q_U p_V}{\mu},
\ee
 where $T^{UU}_{(+)}= g^{UV} g^{UV} T_{VV}^{(+)}$. To get rid of  the volume factor vol($\mathbb{H}_{d-1}$) we write $p_V$ in terms of the total momentum of the signal, which is $p_V^\text{tot}= \int d^{d+1}x \sqrt{-g} T_{VV}= \text{vol}(\mathbb{H}_{d-1})\, p_V$. Therefore $p_V = \frac{p_V^\text{tot}}{\text{vol}(\mathbb{H}_{d-1})}$, and we can write
 \be
 \delta =  \frac{4 \pi G_N}{r_0^{d-1}} \frac{q_U p^\text{tot}_V}{d-1}\,.
 \ee 
 The probe approximation $\delta \lesssim 1$ then becomes
 \be \label{eq-probehom}
  p^\text{tot}_V \lesssim \frac{(d-1)r_0^{d-1}}{4 \pi G_N q_U}.
  \ee
The null shift $\Delta V$ can be written in terms of ANE as (\ref{eq-deltav-ANEC}):
 \be\label{4.18}
 \Delta V =\frac{4 \pi G_N}{d-1} \mathcal{A}_{d}^{\infty}(U_0)\,,\,\,\,\,\,\, \mathcal{A}_d^{\infty}(U_0) \equiv \int_{U_0}^{\infty}  T_{UU}^- dU\,.
 \ee
 Thus, we can write the bound on information transfer by using (\ref{eq-Npvdeltav}), (\ref{eq-probehom}) and (\ref{4.18})
 \be
 N_\text{bits} \lesssim p_V^\text{tot} |\Delta V| \lesssim \left(\frac{(d-1)\, r_0^{d-1}}{4 \pi G_N q_U} \right) \frac{4 \pi G_N}{d-1} |\mathcal{A}_d^{\infty}(U_0)|. 
 \ee

 For homogeneous perturbations, $\mathcal{A}_d^{\infty}(\Delta,U_0)$ is given by (\ref{ANEC_dUTUU}). Note that $|\mathcal{A}_d^{\infty}(\Delta,U_0)| \propto h\,K$, where the proportionality factor is a small number that decreases as we increase $d$, see Fig.~\ref{fig:ANEC}. 
 To investigate in more detail the behavior of the bound on ANE when we increase the spacetime dimensionality, we look into the minimum value of the ANE in (\ref{ANEC_dUTUU}) in terms of $d$. For $t_0=0$, the ANE takes its minimum value at $\Delta = \frac{d}{2}$, and the scaling behavior with respect to $d$ is roughly $\frac{1}{d-a}$ with $a \sim 1.7547 $, which can be seen from Figure \ref{fig:ANEC}. This means that $|\mathcal{A}_d^{\infty}(\Delta,U_0)| \lesssim \frac{1}{d-a}$ with some constant $a$, which shows that the violation of ANEC decreases as we increase $d$. This bounds the information transferred through the wormhole as follows:
 \be
 N_\text{bits} \lesssim \frac{ r_0^{d-1}\,h\, K}{(d-a)  q_U}\,.
 \ee 
 
 This bound qualitatively agrees with the one derived in \cite{Maldacena_2017} in the context of JT gravity when one takes $q_U \sim \mathcal{O}(1)$. It suggests that we can send an arbitrarily large amount of information through the wormhole by increasing $K$. However, this is not the case. The authors of \cite{Freivogel_2020} showed that the above construction only applies for $K \lesssim \frac{1}{G_N}$, which implies
 \be \label{eq-boundNbitd}
  N_\text{bits} \lesssim  \frac{h}{d-a} \,S_\text{BH}\,,
 \ee 
 where we set $q_U=1$. This is in qualitative agreement with the analysis of \cite{Freivogel_2020} for homogeneous shocks, with the difference that (\ref{eq-boundNbitd}) shows how the bound on information transfer scales with $d$. In particular, (\ref{eq-boundNbitd}) shows that the amount of information that can be transferred through the wormhole decreases as we increase the dimensionality of the spacetime. 

\subsubsection*{Backreaction effect} \label{sec-back-info}

\begin{figure}
\begin{center}
\begin{tabular}{cc}
\setlength{\unitlength}{1cm}
\hspace{-0.9cm}
\includegraphics[width=7.5cm]{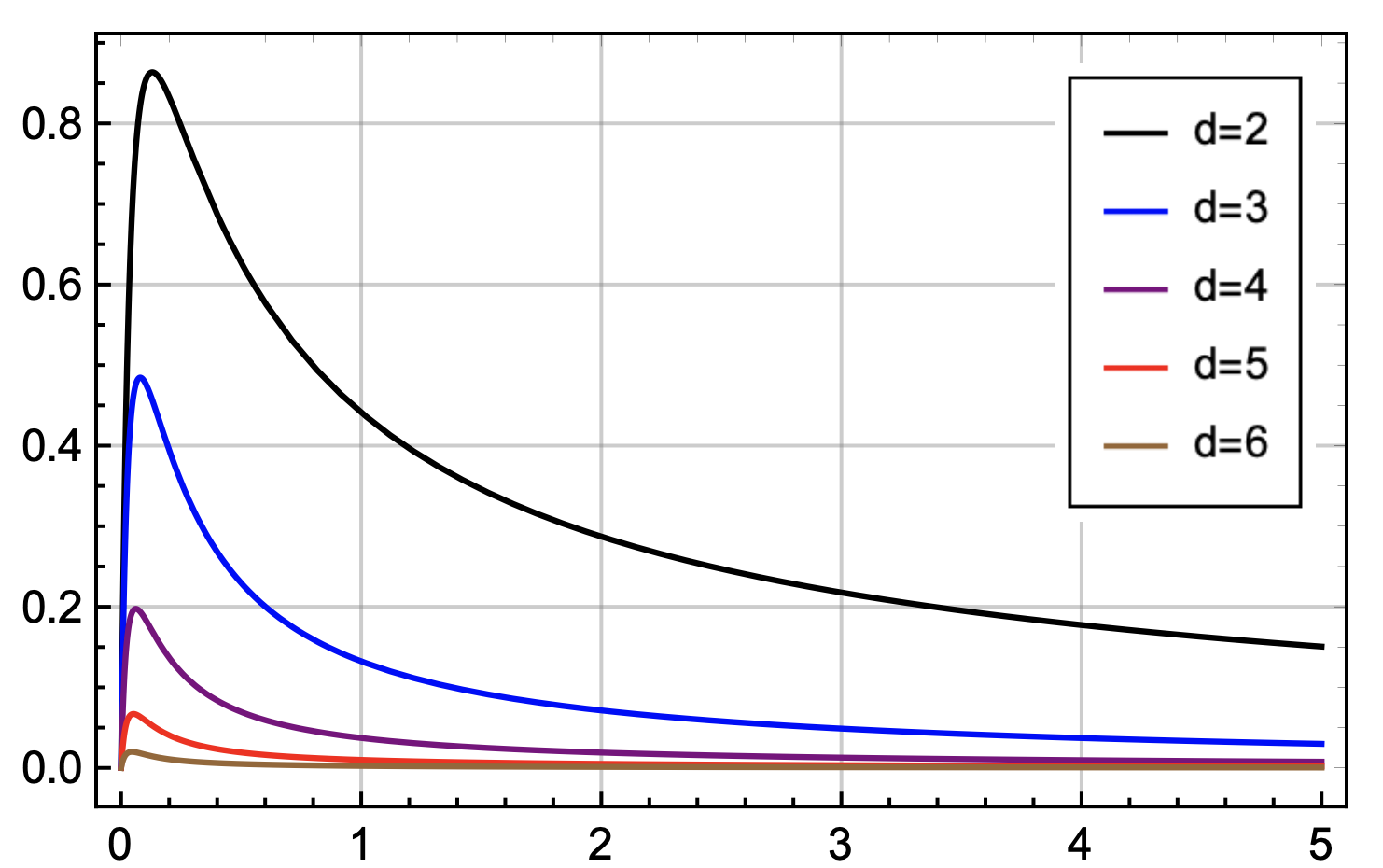}
\qquad \qquad &
\includegraphics[width=7.5cm]{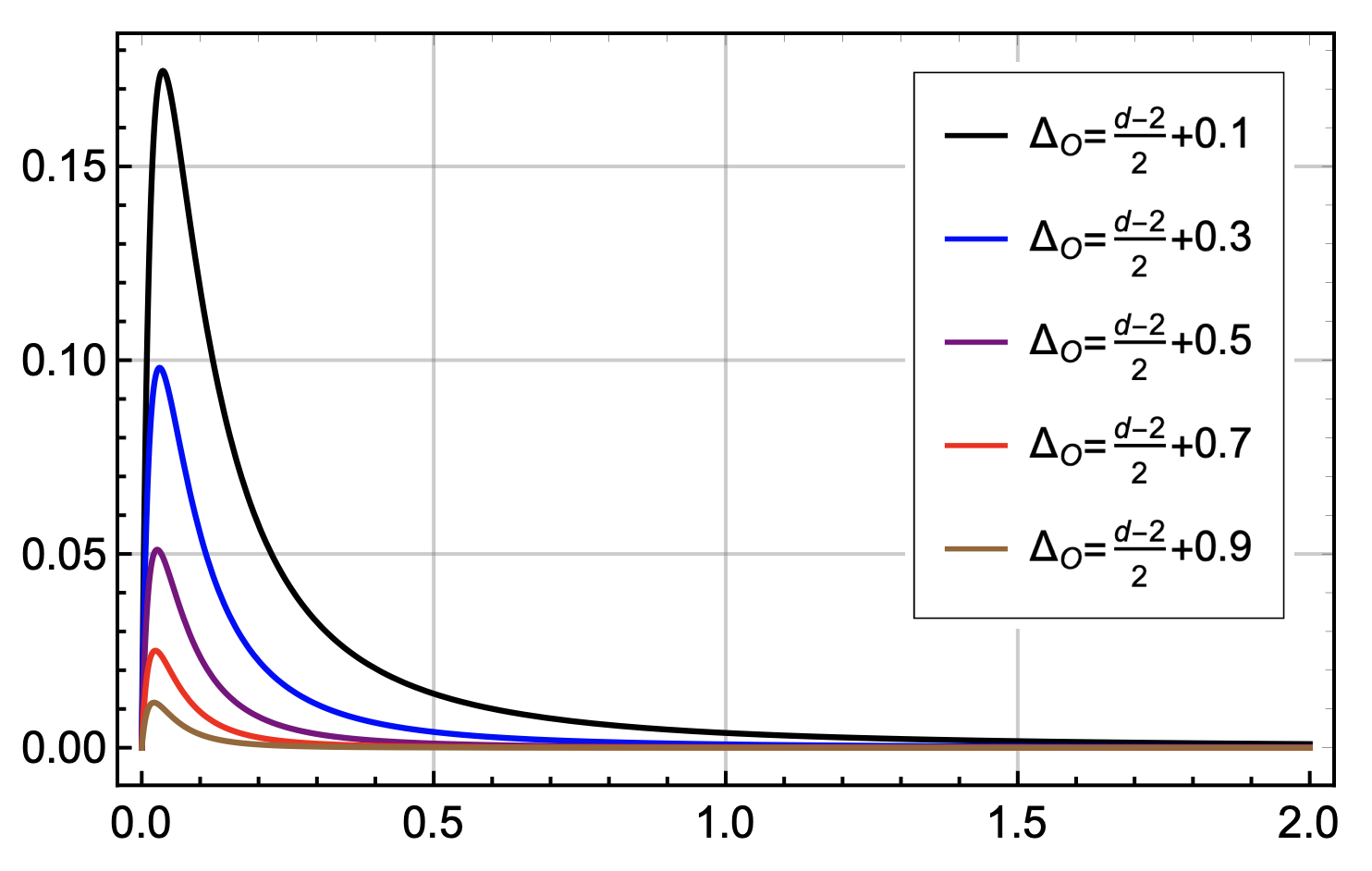}
\qquad
\put(-105,-15){$q^\text{tot}$}
\put(-360,-15){$q^\text{tot}$}
\put(-480,35){\rotatebox{90}{$|\Delta V_\text{back} q^\text{tot}|$}}
\put(-228,35){\rotatebox{90}{$|\Delta V_\text{back} q^\text{tot}|$}}
\end{tabular}
\end{center}
\caption{Bound on information transfer as a function of the total momentum of the signal. On the left panel, we find that the bound on information transfer with backreaction decreases as we increase $d$. Here, we set $\Delta_{\mathcal{O}}=\frac{d-2}{2}+0.5$. On the right panel, we show the dependency on the conformal dimension when the $d$ is fixed. As an example, we draw the $d=3$ case, but other values of $d$ lead to the same qualitative behavior.}
\label{fig-back-info}
\end{figure}

In this section, we briefly discuss the effect of backreaction of the signal on the bound of information transfer. We show in Sec. \ref{sec-backreaction} that the opening of the wormhole decreases as we increase the total momentum of the signal $q^\text{tot}$. As explained in \cite{Caceres_2018}, this implies that the amount of information that can be transferred through the wormhole, which is bounded by  the maximal value of $\Delta V_\text{back} q^\text{tot}$, i.e. 
\be \label{eq-back info}
N \lesssim \text{max}\left[|\Delta V_\text{back} q^\text{tot}| \right]\,.
\ee
In Fig.~\ref{fig-back-info} we plot $\Delta V_\text{back} q^\text{tot}$ versus $q^\text{tot}$ for increasing values of $d$. In this regard, the bound on information transfer decreases as we increase $d$.

 \subsection{Localized shocks} 
We now discuss the case of localized shocks, in which the stress tensor of the signal and of the negative energy read
\be
T_{VV}^{+}=\frac{p_V}{r_{0}^{d-1}}  \delta(V) \delta(\chi-\chi_+), \quad T_{UU}^{-}=\frac{q_U}{r_{0}^{d-1}}  \delta(U) \delta(\chi-\chi_-)\,,
\ee
where $\chi_{\pm}$ denotes the position of the signal/negative energy.
The backreaction of the negative energy reads $ds^{2} \rightarrow ds^{2}+h_{UU}^{-}dU^{2}$, with
\be
h_{UU}^{-}=16 \pi G_{N} \frac{q_U}{r_{0}^{d-1}} \delta (U) f( \chi-\chi_-)\,,
\ee
where $f(\chi) \sim \frac{e^{-(d-1)\chi}}{d}$. The phase shift reads
\be
\delta = \int d^{d+1}x \sqrt{-g} \,T^{UU}_{+} \, h_{UU}^{-}= \frac{4 \pi G_N}{r_0^{d-1}} p_V q_U f(\chi_+ -\chi_{-})\,.
\ee
The probe approximation $\delta \lesssim 1$ then implies
\be
p_V \lesssim \frac{d r_0^{d-1}}{4 \pi G_N q_U}\,,
\ee
where we consider the minimal value of $1/f(\chi_+ -\chi_{-})$.
The final piece of information that we need to find a parametric bound on the information transfer is $\Delta V$, since $N_\text{bits} \lesssim \Delta V p_V$.

\begin{figure}
\begin{center}
\begin{tabular}{cc}
\setlength{\unitlength}{1cm}
\hspace{-0.9cm}
\includegraphics[width=7.5cm]{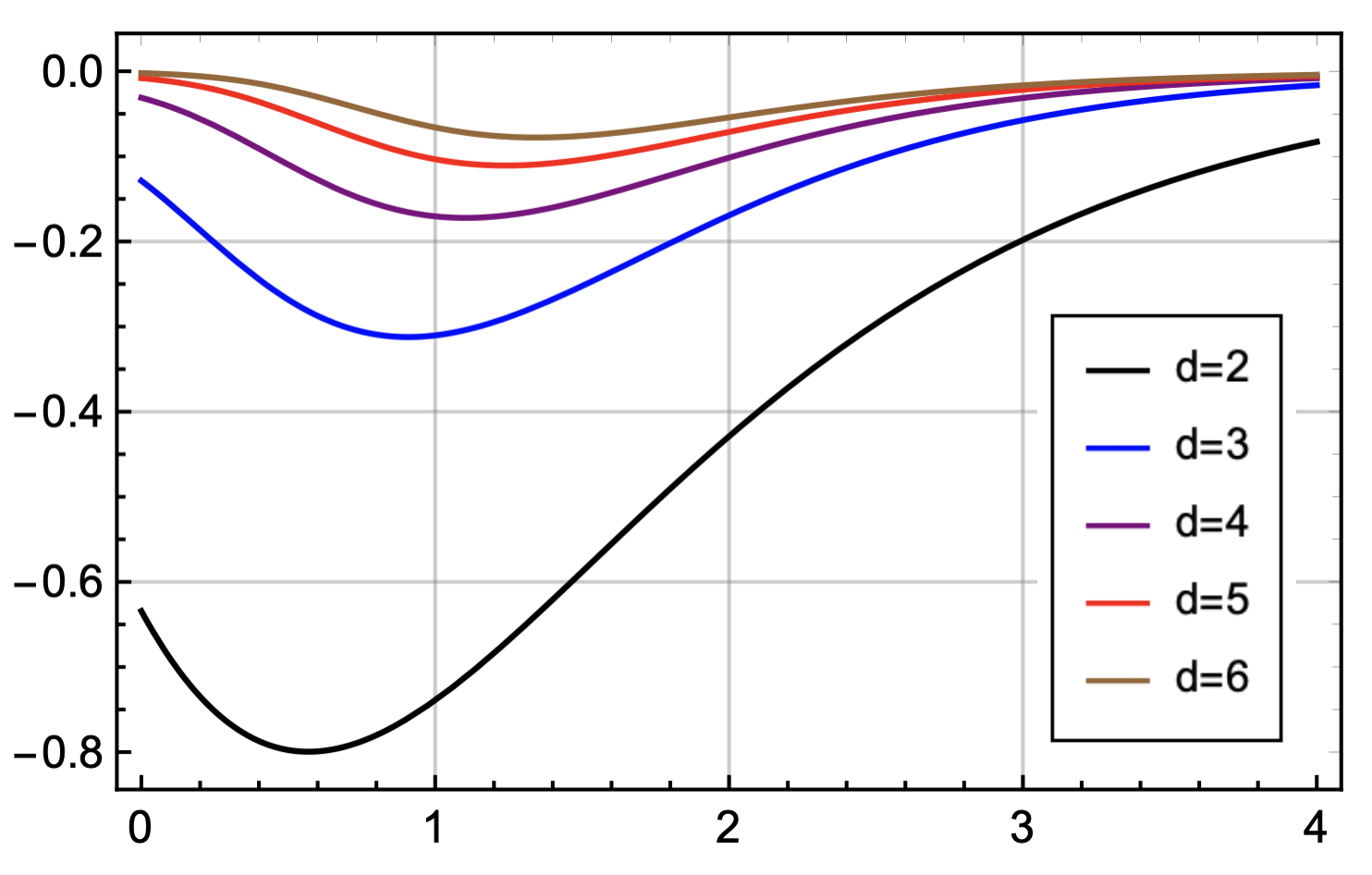}
\qquad\qquad &
\includegraphics[width=7.65cm]{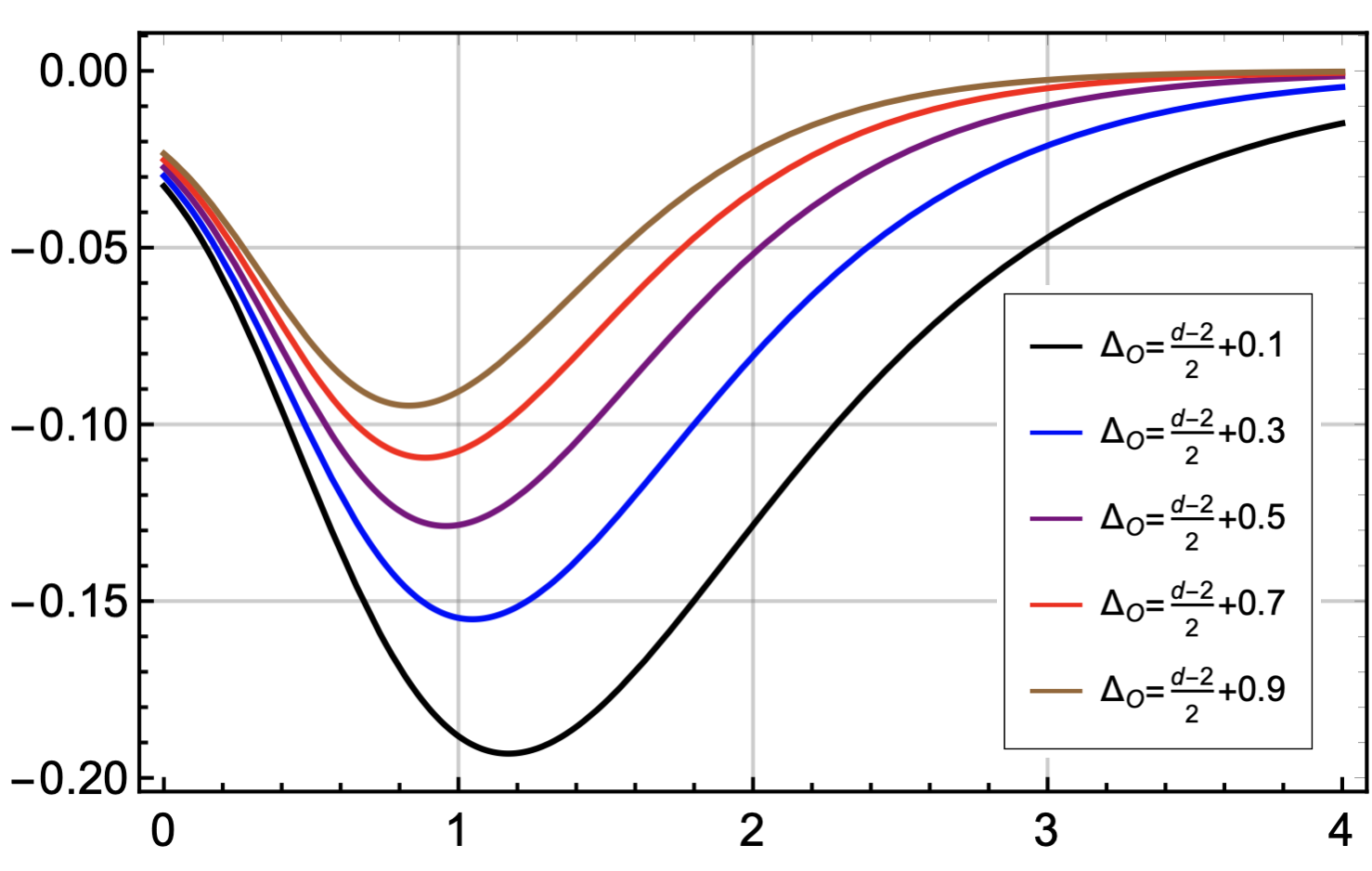}
\qquad
\put(-103,-10){$\chi_{1}$}
\put(-360,-10){$\chi_{1}$}
\put(-485,50){\rotatebox{90}{$\Delta V/C_{D_{1}}$}}
\put(-235,50){\rotatebox{90}{$\Delta V/C_{D_{1}}$}}
\end{tabular}
\end{center}
\caption{ANE for localized shocks as a function of the transverse coordinate of signal $\chi_{1}$ for several values of $d$ (left panel) and for several values of $\Delta_{\mathcal{O}}$ (right panel). We plotted (\ref{eq-deltax1}) as $\Delta V/C_{D_{1}}$ where $C_{D_{1}} \equiv \alpha\,h \, b_{\mathcal{O}}^{2} \frac{\Delta_{\mathcal{O}} \Gamma(2\Delta_{\mathcal{O}})}{4} \text{vol}(S_{d-2})$. We fixed $\Delta_{\mathcal{O}}=\frac{d-2}{2}+0.2$ for each $d$ in the left panel, and we fixed $d=4$ in the right panel. The minimum of $\Delta V$ decreases as we increase $d$.}
\label{fig-loc}
\end{figure}

In the case of local shocks, we have
\be \label{eq-deltax1}
\Delta V(\chi_1) = 16 \pi G_N \, h \, K \, C_{D_{1}} \, \int_{0}^{\infty} d \chi_{4} \frac{\sinh^{d-2} \chi_4}{( \cosh \chi_4 )^{2\Delta+1}} \frac{e^{-(d-1)|\chi_4-\chi_1|}}{d}\,.
\ee
where $C_{D_{1}} \equiv \alpha\,h \, b_{\mathcal{O}}^{2} \frac{\Delta_{\mathcal{O}} \Gamma(2\Delta_{\mathcal{O}})}{4} \text{vol}(S_{d-2})$. The correlator (\ref{eq-Cprobenumerical}) involves an integral over $\chi_1$, so the amount of information transfer varies as we vary $\chi_1$.  The exponential dependence makes $\Delta V$ much smaller than the corresponding quantity in the homogeneous case. This suggests a more constrained bound for local shocks. By numerically studying the behavior of the minimum value of $\Delta V$ in (\ref{eq-deltax1}) as a function of $d$ (see Fig.~\ref{fig-loc}), we find that $\Delta V$ scales as $\frac{1}{d-b}$ with $b \sim 1.1134$. Proceeding as before, we can show that
 \be \label{eq-boundNbitd2}
  N_\text{bits} \lesssim  \frac{h}{d-b} \,S_\text{BH}\,,
 \ee 
which implies that the bound on information transfer in the case of localized shocks also decreases as we increase the dimensionality of the spacetime. We note that the bound is saturated only when the transverse bulk position of the signal matches the transverse bulk position of the negative energy shock wave. This suggests that in practice the number of bits that can be transferred in the case of localized shocks is actully much smaller than (\ref{eq-boundNbitd2}) suggests.

We should note that not all this information reaches the right boundary of the geometry. This is because local perturbations have non-zero angular momentum quantum numbers that introduce a potential barrier that tends to keep them in the near horizon region \cite{Freivogel_2020}. It would be interesting to  follow the approach used in \cite{Freivogel_2020} to investigate this effect in our setup and possibly derive sharper bounds in the case of localized shocks.

Lastly, we observe that the decrease of the ANEC violation for $d \geq 2$ can be explained by the fact that the ANE scales roughly as $\frac{1}{2\Delta_\mathcal{O}+1}$ (see (\ref{ANEC_dUTUU})), and the range of values that we choose for $\Delta_\mathcal{O}$ depends on $d$ as $\frac{d-2}{2} < \Delta <\frac{d}{2}$. Combining these facts we can see that the ANE scales roughly as $\frac{1}{d-a}$, where $a$ is some constant.


 \section{Change of entropy} \label{sec-deltaE}

In this section, we compute the change of energy of the CFT state that results from the double trace deformation, and the corresponding change of entropy.
 Let's focus on the right boundary with Hamiltonian $H_{R}$. Then the expectation value $\langle \psi(t)| H_{R}|\psi(t)\rangle$ can be derived with the state
\be
|\psi \rangle = e^{-i H_{0}(t-t_{0})} U(t,t_{0})|\text{TFD}\rangle
\ee
\be
U(t,t_{0})=\mathcal{T} e^{-i\int_{t_{0}}^{t} d t_1 \delta H(t_1)}
\ee
where thermofield double state is defined as (\ref{eq-tfd}). By considering the result at first order in $h(t_1)$, we can write
\be
\begin{split} \label{eq-energy change}
    \delta E_{R} &=  i \int_{t_{0}}^{t} dt_{1} \int d{\bf x} \, h(t_{1})\,  \langle \text{TFD}|\, [\delta H(t_1),H_R]\, |\text{TFD} \rangle \\
    & = \int_{t_{0}}^{t} dt_{1} \int d{\bf x} \, h(t_{1})\,  \langle \text{TFD}|\, \partial_{t} \mathcal{O}_{R}(t_1,{\bf x}) \mathcal{O}_{L}(-t_1,{\bf x}) |\text{TFD} \rangle\\
    & = h \left( \frac{2 \pi}{\beta} \right) \int_{t_{0}}^{t} d t_{1}\int d{\bf x}  \,\langle \text{TFD}|  \, \partial_{t} \mathcal{O}_{R}(2t_1+i\beta/2,{\bf x})\mathcal{O}_{R}(0,{\bf x})|\text{TFD} \rangle\\
     & = h \,c_{\Delta}\int d{\bf x} \bigg[\frac{1}{(\cosh(2t)+\cosh( d({\bf x},{\bf x}'))^{\Delta}}-\frac{1}{(\cosh(2 t_{0})+\cosh d({\bf x},{\bf x}'))^{\Delta}}\bigg]\,,
\end{split}
\ee
where $c_{\Delta}$ is given by (\ref{eq-cdelta}).
 In the second line, we have used the KMS condition. Here, if we turn off the interaction at time $t_\text{f}$, $t$ in (\ref{eq-energy change}) is replaced by $t_\text{f}$. 
 Assuming $h>0$, the change of energy is positive for $t_{0} < t_\text{f} < 0$, and negative for $0 < t_0 < t_\text{f}$, and the wormhole is traversable in both cases. This feature was also observed in lower dimensional cases ($d \leq 2$)~\cite{almheiri2018escaping,Bak_2018}.

We find a closed form for the change of energy by directly integrating (\ref{eq-energy change}). The result can be written in terms of the Appell hypergeometric function 
\begin{multline} \label{eq-deltaE}
\delta E_{R}= h\, c_{\Delta} \, \text{vol}(S_{d-2}) \frac{\Gamma(\Delta-d+2)\Gamma(\frac{d-1}{2})}{\Gamma(\Delta-\frac{d-3}{2})}\bigg[F_{1}\bigg(\Delta-d+2;\Delta,\frac{3-d}{2};\Delta-\frac{d-3}{2};-\cosh(2  t_\text{f}),-1\bigg)\\
-F_{1}\bigg(\Delta-d+2;\Delta,\frac{3-d}{2};\Delta-\frac{d-3}{2};-\cosh(2 t_{0}),-1\bigg)\bigg]\,,
\end{multline}
where we consider operators that do not depend on the angles on $S_{d-2}$. The above formula is valid for $\Delta > 2-d$ and $d-1>0$, which is always true for the cases we consider, in which $d \geq 2$ and $\Delta \geq d/2$.

We can now use the first law of entanglement to compute the change of entropy as $\delta S= T \delta E_R = \frac{1}{2\pi} \delta E_R$. This implies that the entropy of the topological black hole, or, equivalently, the entanglement entropy between the two sides of the geometry, reduces for $0 < t_0 < t_\text{f}$, since in this case $\delta E_R <0$. As explained in \cite{Bak_2018}, from the boundary perspective the decrease of entropy can be viewed as a result of the measurement in quantum teleportation. This fact can be used to interpret the information transfer through the wormhole as quantum teleportation. 

Finally, it has been pointed out in \cite{Gao_2017} that the change of entropy $\delta S$ resulting from the double trace deformation can in principle be computed in the bulk in terms of quantum extremal surfaces~\cite{Engelhardt_2015}. It would be interesting to investigate that by following the ideas presented in \cite{chen2020quantum}.

\section{Discussion} \label{sec-disc}

In this work, we have studied the Gao-Jafferis-Wall (GJW) holographic teleportation protocol in a higher dimensional setup ($d \geq 2$). In particular, we consider the hyperbolic slicing of a pure AdS geometry, which can be thought of as a topological hyperbolic black hole. The maximally extended geometry contains two exterior regions, the Rindler wedges of AdS, which are connected by a wormhole. 

We show that a double trace deformation involving a non-local interaction between operators in the left and right boundaries can make the above-mentioned wormhole traversable, allowing a sign to be transmitted between the two Rinlder wedges of AdS. The traversability is due to a violation of the average null energy condition (ANEC) in the bulk. We compute the average null energy using two different methods: the point splitting method of GJW \cite{Gao_2017}, and the eikonal method used in \cite{Maldacena_2017}. We generalize both methods to our higher dimensional setup and show that they give consistent results. In particular, we find an analytic formula for the ANE that nicely generalizes GJW result to higher dimensions ($d \geq 2$), and reduces to GJW result when we set $d=2$. See (\ref{ANEC_dUTUU}). In particular, we show in Fig.~\ref{fig:ANEC} that once we fix the coupling strength of the deformation the violation of ANEC reduces very quickly as we increase the dimensionality of the spacetime.

Our setup evades the no-go theorem derived in \cite{Freivogel_2019}, which establishes that semiclassical eternal traversable wormholes are not possible in spacetime dimensions higher than two. This theorem is derived under the assumption of Poincare invariance in the boundary direction, and with the use of Weyl invariant matter fields. Moreover, in the setup of \cite{Freivogel_2019} the two boundary CFTs are not entangled with each other. Our setup is different for a few reasons: (i) our traversable wormhole is not eternal, i.e., the wormhole becomes only becomes traversable after we introduce the double trace deformation at some time $t_0$, (ii) the two boundary CFTs in our setup are entangled in a thermofield double state, (iii) the matter fields we use do not have Weyl symmetry. Therefore, our result is not in contradiction with \cite{Freivogel_2019}.


The Rindler-AdS geometry allows us to find several other analytic results, including closed formulas that account for effects of backreaction (\ref{eq-deltavback2}), the change of energy of the CFT state (\ref{eq-deltaE}), and two-sided correlation functions that diagnose traversability (\ref{eq-Cprobenumerical}). 

We checked that the optimal condition for traversability is determined by the butterfly speed $v_B$. In fact, we show in Sec.~\ref{sec-localized} that the sweet spot for traversability moves with the butterfly speed. This is in accordance with previously established results \cite{Couch_2020}, but our setup provides the first example in which the butterfly cone is distinguishable from the light-cone, i.e., the sweet spot moves with $v_B <1$.

In Sec.~\ref{sec-bound}, we derive parametric bounds on information transfer and discuss how these bounds are affected by the dimensionality of the spacetime. In particular, we show that the information transfer is bounded by the black hole entropy, as observed previously in the literature~\cite{Freivogel_2020}. In the case of homogeneous shocks, the bound on information transfer scales as $\frac{1}{d-a}$ with $a \sim 1.7547$, as seen in Fig.~\ref{fig:ANEC}. For local perturbations, we numerically estimate, based on the results of Fig.~\ref{fig-loc}, that the bound on information transfer scales as $\frac{1}{d-b}$ with $b \sim 1.1134$. The above results are valid for a Rindler-AdS$_{d+1}$ geometry, but the feature that the bound on information transfer reduces as we increase $d$ might be a general feature of higher dimensional systems.



Finally, the change of entropy that results from making the wormhole traversable could in principle be computed using quantum extremal surfaces~\cite{Engelhardt_2015}. It would be interesting to
compute this change of entropy using quantum extremal surfaces and compare the results with our result in Sec.~\ref{sec-deltaE}.
\acknowledgments
We would like to thank Kyung-Sun Lee and Juan Pedraza for helpful discussions. This work was supported by Basic Science Research Program through the National Research Foundation of Korea(NRF) funded by the Ministry of Science, ICT \& Future
Planning (NRF-2017R1A2B4004810) and the GIST Research Institute(GRI) grant funded by the GIST in 2020. 
V. Jahnke was supported by Basic Science Research Program through the National Research Foundation of Korea(NRF) funded by the Ministry of Education(NRF-2020R1I1A1A01073135). 
B. Ahn was also supported by Basic Science Research Program through the National Research Foundation of Korea funded by the Ministry of Education (NRF-2020R1A6A3A01095962).

\appendix 
\section{Details about the ANE calculation} \label{app:A}
In this section, we derive (\ref{ANEC_dUTUU}) in detail using the point splitting method. We first use (\ref{eq-stress}) to write:
\begin{align}\label{eq-ANEC}
\mathcal{A} \equiv  \int_{U_0}^{\infty}  T_{UU} dU =& \kappa_d \int_{U_0}^{\infty} dU \lim_{U'\rightarrow U} \partial_U G(U,U';U_0) \nonumber \\
=& \kappa_d \bigg[ G(\infty,\infty;U_0)- G(U_0,U_0;U_0) -\int_{U_0}^{\infty} dU \partial_U^{(2)} G(U,U;U_0)  \bigg]
\end{align}
where
\be
\kappa_d =  -4 \Delta h\, \text{vol}(S_{d-2})\, 	\frac{2 \sin{\pi \Delta}}{\pi^d (2\Delta-d)^2} \bigg(\frac{\Gamma(\Delta)}{2^\Delta \Gamma (\Delta - \frac{d}{2})} \bigg)^2
\ee
and 
\be
G(U,U';U_0) \equiv \int_{U_0}^U dU_1 \int_{1}^{U/U_1} \frac{dy}{ (y^2-1)^{\frac{3-d}{2}}}\frac{U_1^\Delta}{(U-U_1 y)^\Delta (U' U_1+y)^{\Delta+1}} \label{GUU} 
\ee
For $d=2$, GJW~\cite{Gao_2017} showed that $G(\infty,\infty;U_0)=0$ and $G(U_0,U_0;U_0)=0$ for $\Delta < \frac{3}{2}$ so that only the last term of the \eqref{eq-ANEC} contributes to the average null energy. We show in the following that this is also true in our case. The integral with respect to $y$ in \eqref{GUU} can be written as an Appell hypergeometric function
\begin{align} \label{eq-Iy}
I_y= & \int_{1}^{U/U_1} \frac{dy}{(y^2-1)^{\frac{3-d}{2}}}\frac{U_1^\Delta}{(U-U_1 y)^\Delta (U' U_1+y)^{\Delta+1}}\\
=&\frac{\Gamma(\frac{d-1}{2}) \Gamma(1-\Delta)}{\Gamma(-\Delta+\frac{d+1}{2})} \frac{2^{\frac{d-3}{2}}  }{(\frac{U}{U_1}-1)^{\Delta-\frac{d-1}{2}} (U' U_1+1)^{\Delta+1}}  \nonumber\\
&F_1\bigg( \frac{d-1}{2};\frac{3-d}{2},\Delta+1 ; -\Delta+ \frac{d+1}{2};\frac{U_1-U}{2U_1},\frac{U_1-U}{U_1(1+U'U_1)}    \bigg) \,.\nonumber
\end{align}
Here, we have used the integral form of Appell hypergeometric function 
\be \label{eq-Appel}
F_{1}(a;b_{1},b_{2};c;x,y)=\frac{\Gamma(c)}{\Gamma(a)\Gamma(c-a)} \int_{0}^{1} dt \,\,\, t^{a-1} (1-t)^{c-a-1}(1-xt)^{-b_{1}}(1-yx)^{-b_{2}}
\ee
with $x=\frac{y-1}{U/U_{1}-1}$.
The above formula is valid for $\text{Re}(a) >$ and $\text{Re}(c)>0$. We now show that both $G(\infty,\infty;U_0)$ and $G(U_0,U_0;U_0)$ vanish, which implies that $\int dU T_{UU}$ can be obtained from the last term in \eqref{eq-ANEC}. We first use (\ref{eq-Iy}) to rewrite (\ref{GUU}) as
\begin{equation}
G(U,U';U_{0})=\kappa_d  \frac{\Gamma(\frac{d-1}{2}) \Gamma(1-\Delta)}{2^{\frac{3-d}{2}} \Gamma(\frac{d+1}{2}-\Delta)} \int_{U_{0}}^{U} dU_{1} \frac{F_1\bigg( \frac{d-1}{2};\frac{3-d}{2},\Delta+1 ; -\Delta+ \frac{d+1}{2};\frac{U_1-U}{2U_1},\frac{U_1-U}{U_1(1+U'U_1)} \bigg) }{(\frac{U}{U_1}-1)^{\Delta-\frac{d-1}{2}} (U' U_1+1)^{\Delta+1}}
\end{equation}
By defining $z=\frac{U_{1}-U_{0}}{U-U_{0}}$, we can write $G(U_0,U_0;U_0)$ as
\be
G(U_{0},U_{0};U_{0}) \propto \lim_{U\rightarrow U_{0}} \int_{0}^{1} dz (U-U_{0})^{\frac{d+2}{2}-\Delta}\frac{U_{0}^{\Delta-\frac{d-1}{2}}}{(1-z)^{\Delta-\frac{d-1}{2}}} F_{1}\bigg(\frac{d-1}{2};\frac{3-d}{2},\Delta+1 ; -\Delta+ \frac{d+1}{2};0,0\bigg)
\ee
which vanishes for $\Delta < \frac{d+1}{2}$. Next, we can replace $U$ and $U'$ by $\infty$ in order to get the first term in (\ref{eq-ANEC}):
\be
G(\infty,\infty;U_{0}) \propto \lim_{U\rightarrow \infty}\frac{1}{U^{2\Delta+1}}\int_{0}^{1} dz \frac{(1+z)^{\frac{d-3}{2}}(1-z)^{\frac{d-1}{2}-\Delta}}{z^{d-1}} {}_{2} F_{1} \bigg(\frac{3-d}{2},1-\Delta,\frac{d+1}{2}-\Delta,\frac{1-z}{1+z} \bigg)\,.
\ee
Using the above expression, we numerically checked that $G(\infty,\infty;U_{0})$ also vanishes. Collecting these results, we can write
\bea
\mathcal{A}&=&\int_{U_{0}}^{\infty} dU \lim_{U'\rightarrow U} \partial_{U'} \int_{U_{0}}^{U} dU_{1} \int_{1}^{U/U_{1}} dy (y^{2}-1)^{\frac{d-3}{2}}\frac{U_{1}^{\Delta}}{(U-U_{1}y)^{\Delta}(y+U_{1}U')^{\Delta+1}}\\
&=& -\int_{U_{0}}^{\infty} dU \int_{U_{0}}^{U} dU_{1} \int_{1}^{U/U_{1}} dy (y^{2}-1)^{\frac{d-3}{2}} \label{eq-A10} \frac{(\Delta+1)U_{1}^{\Delta+1}}{(U-U_{1}y)^{\Delta}(y+U_{1}U)^{\Delta+2}}\\
&=& -\int_{U_{0}}^{\infty} dU_{1} \int_{U_{1}}^{\infty} dU \int_{1}^{U/U_{1}} dy (y^{2}-1)^{\frac{d-3}{2}} \frac{(\Delta+1)U_{1}^{\Delta+1}}{(U-U_{1}y)^{\Delta}(y+U_{1}U)^{\Delta+2}}\\
&=& -\int_{U_{0}}^{\infty} dU_{1} \int_{U_{1}}^{\infty} dU \,\,\, \mathcal{I}_{y}
\eea
In the third line, we have changed the limits of the first two integrals while covering the same region of integration. By changing variables as $z=\frac{U-U_{1}y}{U-U_{1}}$, we can express $\mathcal{I}_{y}$ in terms of an Appell hypergeometric function
\bea \label{eq-constrained hyper}
\mathcal{I}_{y}&&=\frac{(\Delta+1)(U+U_{1})^{\frac{d-3}{2}}U_{1}^{2\Delta+5-d}}{(1+U_{1}^{2})^{\Delta+2}(U-U_{1})^{\Delta-\frac{d-1}{2}}U^{\Delta+2}}\frac{\Gamma(1-\Delta)\Gamma(\frac{d-1}{2})}{\Gamma(\frac{d+1}{2}-\Delta)}\\
&& \times F_{1}\bigg(1-\Delta;\frac{3-d}{2},\Delta+2;\frac{d+1}{2}-\Delta,\frac{U-U_{1}}{U+U_{1}};\frac{U-U_{1}}{U(1+U_{1}^{2})} \bigg)\,.
\eea
Here, we used the integral form  (\ref{eq-Appel}) of the Appell hypergeometric function. The constraints $\text{Re}(c) >0$ and $\text{Re}(a)>0$ imply $\frac{d+1}{2}-\Delta >0$ and $\Delta<1$\footnote{Despite being an essential condition for the formula (\ref{eq-constrained hyper}) to be valid, the condition $\Delta<1$ does not seem to be a necessary condition for our final formula (\ref{ANEC_dUTUU}) for $\int dU T_{UU}$, which seems to be valid at least for $\Delta \leq \frac{d+1}{2}$. In fact, (\ref{ANEC_dUTUU}) can also be derived using the eikonal approximation without any constraint on the upper value of $\Delta$.}. Next, we consider the integral in $U$. Changing variables as $\omega=\frac{U-U_{1}}{U+U_{1}}$, we write
\be \label{eq-ANEC dU dU1}
\mathcal{A}=-\frac{\kappa_{\Delta}}{2^{\Delta-d+1}} \sum_{m} \frac{(1-\Delta)_{m}(\Delta+2)_{m}}{(\frac{d+1}{2}-\Delta)_{m} m!} 2^{m} \int_{U_{0}}^{\infty} dU_{1} \frac{U_{1}^{2}}{(1+U_{1}^{2})^{\Delta+2+m}} \mathcal{I}_{\omega}\,,
\ee
where
\be
\mathcal{I}_{\omega}=\int_{0}^{1} d \omega \frac{\omega^{m-\Delta+\frac{d-1}{2}(1-\omega)^{2\Delta+2-d}}}{(1+\omega)^{m+\Delta+2}} {}_{2} F_{1} \bigg( 1-\Delta+m,\frac{3-d}{2};\frac{d+1}{2}-\Delta+m;\omega \bigg)\,.
\ee
To derive the above formula, we used the following relation between $F_1$ and ${}_2F_1$:
\be
F_{1}\big(a;b_{1},b_{2};c;x,y \big)=\sum_{m} \frac{(a)_{m}(b_{2})_{m}}{(c)_{m}} \frac{y^{m}}{m!} {}_{2} F_{1}\big(a+m,b_{1};c+m;x \big)
\ee
where $(q)_{m}=\frac{\Gamma(q+m)}{\Gamma(q)}$ is the rising Pochhammer symbol. By some manipulation, $\mathcal{I}_{\omega}$ becomes
\be \label{eq-ANEC dU}
\mathcal{I}_{\omega}=\frac{1}{2^{\Delta+2+m}} \frac{\Gamma(m-\Delta+\frac{d+1}{2})\Gamma(2\Delta+3-d)\Gamma(2\Delta+1)}{\Gamma(2\Delta+\frac{5-d}{2})\Gamma(\Delta+2+m)} {}_{2} F_{1}\bigg(2\Delta+3-d,2\Delta+1;2\Delta+\frac{5-d}{2};\frac{1}{2}\bigg)\,.
\ee
For the above result, we used the identities:
\bea
&&\int_{0}^{y} d x \frac{x^{c-1}(y-x)^{\beta-1}}{(1-z x)^{\rho}} {}_{2} F_{1} \left(a,b;c;\frac{x}{y} \right)=
\frac{y^{c+\beta-1}}{(1-y z)^{\rho}} \frac{\Gamma(c) \Gamma(\beta) \Gamma(c-a-b+\beta)}{\Gamma(c-a+\beta)\Gamma(c-b+\beta)}\\
&&\times {}_{3} F_{2}\left(\beta,\rho,c-a-b+\beta;c-a+\beta,c-b+\beta;\frac{y z}{y z-1}\right),
\eea
and 
\be
{}_{3} F_{2} \left(a_{1},a_{2},a_{3};b_{1},a_{2};z \right) = {}_{2} F_{1} \left(a_{1},a_{3};b_{1};z \right)
\ee
Lastly, we consider the integral in $U_{1}$ in (\ref{eq-ANEC dU}):
\be \label{eq-ANEC dU1}
\int_{U_{0}}^{\infty} d U_{1} \frac{U_{1}^{2}}{(1+U_{1}^{2})^{\Delta+2-m}}=\frac{1}{2(m+\Delta+\frac{1}{2})} {}_{2}F_{1}\left( m+\Delta+\frac{1}{2},-\frac{1}{2};m+\Delta+\frac{3}{2};\frac{1}{1+U_{0}^{2}}\right)
\ee
Here, we used
\be
{}_{2}F_{1}\left(a,b;c;z\right)=(1-z)^{-a} {}_{2}F_{1}\left( a,c-b;c;\frac{z}{z-1}\right)\,.
\ee
Then we get the following result
\bea
\mathcal{A}&&=-\frac{2^{d-4-2\Delta}\kappa_{\Delta}}{(1+U_{0}^{2})^{\Delta+\frac{1}{2}}}\left[\frac{\Gamma(\frac{1}{2})\Gamma(2\Delta+\frac{5-d}{2})}{\Gamma(\Delta+\frac{4-d}{2})\Gamma(\Delta+1))}\right] \frac{\Gamma(2\Delta+3-d)\Gamma(2\Delta+1)}{\Gamma(2\Delta+\frac{5-d}{2})} \nonumber\\
&&\sum_{m} \frac{(1-\Delta)_{m}(\Delta+2)_{m}}{(\frac{d+1}{2}-\Delta)_{m}m!}\frac{(\Delta+\frac{1}{2})_{m}}{(\Delta+\frac{3}{2})_{m}}\frac{1}{(1+U_{0}^{2})^{m}} {}_{2} F_{1}\bigg(\Delta+\frac{1}{2}+m,-\frac{1}{2};\Delta+\frac{3}{2}+m;\frac{1}{1+U_{0}^{2}}\bigg) \nonumber\\
\eea
where we plugged (\ref{eq-ANEC dU}), (\ref{eq-ANEC dU1}) into (\ref{eq-ANEC dU dU1}) and used the relation
\be
{}_{2} F_{1}\big(a,b;\frac{a+b+1}{2};\frac{1}{2} \big)=\frac{\Gamma(\frac{1}{2})\Gamma(\frac{1+a+b}{2})}{\Gamma(\frac{a+1}{2})\Gamma(\frac{b+1}{2})}
\ee
Finally, by using
\bea
&&\sum_{m}\frac{(1-\Delta)_{m}(\Delta+\frac{1}{2})_{m}}{m! (\Delta+\frac{3}{2})_{m}}\left(\frac{1}{1+U_{0}^{2}}\right)^{m} {}_{2}F_{1}\left(\Delta+\frac{1}{2}+m,-\frac{1}{2},\Delta+\frac{3}{2}+m;\frac{1}{1+U_{0}^{2}}\right) \nonumber\\
&&={}_{2}F_{1}\left(\Delta+\frac{1}{2},\frac{1}{2}-\Delta,\Delta+\frac{3}{2};\frac{1}{1+U_{0}^{2}}\right),
\eea
and some simple manipulations obtain a closed form for the ANE:
\bea
&&\frac{1}{\text{vol}(S_{d-2})}\int T_{UU}dU\\
&&= - \frac{h \pi^{\frac{1}{2}-d} \Gamma(\frac{d-1}{2} )}{(2\Delta +1)}\frac{\Gamma(\Delta+\frac{1}{2}) \Gamma(\Delta+ \frac{3-d}{2})}{\Gamma(\Delta+1-\frac{d}{2} )^2}
 \frac{{}_2F_1\bigg(\Delta+\half, \half-\Delta, \Delta+\frac{3}{2};\frac{1}{1+U_0^2}\bigg)
}{ (1+U_0^2)^{\Delta+\frac{1}{2}}}\,,
\eea
which is valid for a Rindler-AdS$_{d+1}$ geometry deformed by a non-local interaction of the form (\ref{eq-deformation}). The unitarity condition for the scalar operators implies $\Delta > \frac{d}{2}-1$, while the condition for the perturbation to be relevant is $\Delta \leq d/2$. If we do not restrict our calculation to relevant perturbations, our formula seems to be valid for $\Delta < \frac{d+1}{2}$. The same formula can be obtained using the eikonal approximation with (apparently) no upper bound for $\Delta$.

\bibliographystyle{JHEP}

\end{document}